# An element-wise approach for simulating transcranial MRI-guided focused ultrasound thermal ablation

Nathan McDannold[1], P. Jason White[1], and Rees Cosgrove[2]

*[1]Department of Radiology, Brigham and Women's Hospital and Harvard Medical School, Boston, Massachusetts 02115, USA*

*[2]Department of Neurosurgery, Brigham and Women's Hospital and Harvard Medical School, Boston, Massachusetts 02115, USA*

This work explored an element-wise approach to model transcranial MRI-guided focused ultrasound (TcMRgFUS) thermal ablation, a noninvasive approach to neurosurgery. Each element of the phased array transducer was simulated individually and could be simultaneously loaded into computer memory, allowing for rapid calculation of the pressure field for different phase offsets used for beam steering and aberration correction. We simulated the pressure distribution for 431 sonications in 32 patients, applied the phase and magnitude values used during treatment, and estimated the resulting temperature rise. We systematically varied the relationship between CT-derived skull density and the acoustic attenuation and sound speed to obtain the best agreement between the predictions and MR temperature imaging (MRTI). The optimization was validated with simulations of 396 sonications from 40 additional treatments. After optimization, the predicted and measured heating agreed well ($R^2$: 0.74 patients 1-32; 0.71 patients 33-72). The dimensions and obliquity of the heating in the simulated temperature maps correlated well with the MRTI ($R^2$: 0.62, 0.74 respectively), but the measured heating was more spatially diffuse. The energy needed to achieve ablation varied by an order of magnitude (3.3-36.1 kJ). While this element-wise approach requires more computation time up front, it can be performed in parallel. It allows for rapid calculation of the three-dimensional heating at the focus for different phase and magnitude values on the array. We also show how this approach can be used to optimize the relationship between CT-derived skull density and acoustic properties. While the relationships found here need further validation in a larger patient population, these results demonstrate the promise of this approach to model TcMRgFUS.

## I. INTRODUCTION

Transcranial MRI-guided focused ultrasound has emerged as a non-invasive neurosurgical approach for thermal lesioning in the brain. This method uses a hemispherical phased array transducer to correct for aberrations of the acoustic field caused by the skull, allowing for accurate focusing to central regions in the brain. The method is clinically-approved in several countries for thalamotomy for essential tremor [1] and tremor-dominant Parkinson's Disease [2], and is under investigation for a number of other functional neurosurgery applications, including thalamotomy for neuropathic pain [3], pallidotomy for Parkinson's Disease [4], and capsulotomy for obsessive-compulsive disorder [5]. Similar systems are being tested clinically to disrupt the blood-brain barrier as a treatment for Alzheimer's Disease [6] and to enhance delivery of chemotherapy to patients with glioblastoma [7].

TcMRgFUS thermal ablation has several significant limitations. First, the treatment is currently restricted to central locations in the brain. When the transducer is focused at more peripheral targets, less energy can be transmitted through the bone due to large incidence angles between the transducer elements and the skull, leading to overheating. Furthermore, even at central locations, the shape of the focus can become oblique [8], which can put nearby structures at risk for unwanted thermal damage and side effects [9]. The bony properties of the skull vary substantially between patients, and the energy needed to achieve an ablative thermal dose at the focus can be excessive in some patients. As a result, some patients are not candidates for this treatment [10]. Finally, for reasons not fully understood, as sonications are delivered at escalating acoustic energies, it becomes more difficult to heat the tissue at the focus [11]. Being able to improve the focusing through the skull could ameliorate some of the challenges, expand the patient population who are candidates for TcMRgFUS, and perhaps prevent side effects due to off-target heating.

To correct for aberrations induced by the irregular skull, the TcMRgFUS system estimates changes in of the ultrasound field induced by the bone for each element of the transducer. This correction is based on an acoustic model that uses the geometry of the skull and estimates of bone density obtained from CT scans [12,13]. The density is used to predict the acoustic sound speed and attenuation based on calibrations obtained with cadaver skulls [14-16]. Currently, a simplified proprietary acoustic model is used to generate the phase aberration corrections. Several studies have evaluated three-dimensional acoustic models to predict the focal heating [8,16-18]. Those studies have been able to reasonably predict the temperature and shape of the focus and the acoustic energy needed to reach an ablative thermal dose at the focus.

This work took a different approach to the simulation problem. Instead of simulating the entire acoustic field in one large model, we simulated each transducer element separately (FIG. 1). These simulations were rotated and interpolated into a global frame of reference and saved [19]. All the element-wise simulations could be loaded into computer memory simultaneously, enabling rapid iteration of phase and magnitude elements to simulate different correction schemes (or no correction) and perform beam steering to different targets. Ultimately, we aim to compile a library of these small simulations to produce a look-up table and to provide inputs for machine learning, which could enable use to rapidly simulate the heating during a treatment by relating each element to a previously-simulated one. Rapid optimization of the phase and magnitude could maximize the peak intensity at the focus and better define the shape and size of the focal ablative region.

## II. METHODS

### A. TcMRgFUS treatments

The treatments used the ExAblate Neuro transcranial MRI-guided focused ultrasound (TcMRgFUS) system (InSightec, Haifa, Israel), which operated at 660 kHz. This device has a 1024-element hemispherical phased array (993 active elements at our site) integrated with a 3T MRI (GE750, GE Healthcare). The patient was placed in a stereotactic frame which was attached to the MRI table. A flexible membrane was attached to the patient's head and the open face of the transducer. Acoustic coupling was achieved with degassed water that is chilled and circulated between sonications to minimize skull heating. The transducer was attached to a manually-operated positioning system that was steered so that the focus was within ±1 mm of the initial brain target identified by the neurosurgeon. Additional steering from this location was achieved electronically using the phased array. The location of the transducer was found in the MRI space via MRI tracking coils, and imaging was performed using the body coil.

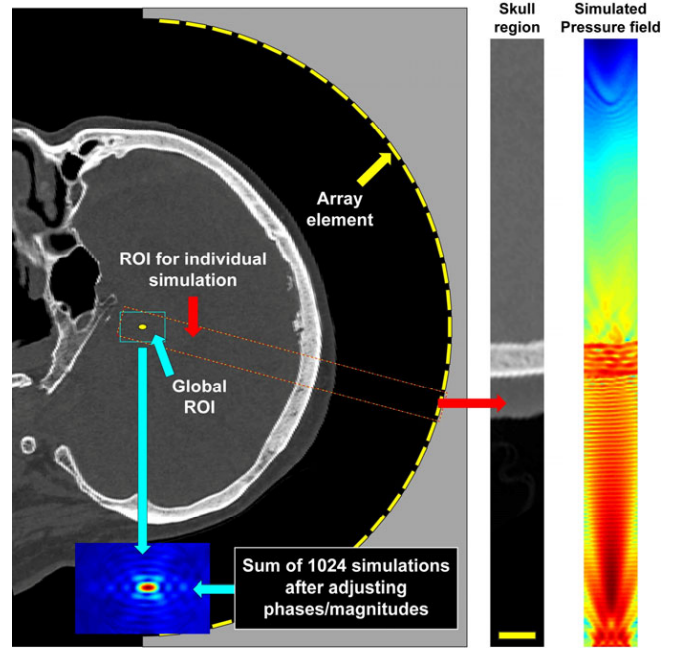

FIG. 1. Methods. Left: Sagittal reformat of a CT scan; a diagram of the 30 cm diameter 650 kHz TcMRgFUS phased array is superimposed. A small volume is selected that includes one array element. Right: The pressure field (log scale shown) is simulated in this volume using the CT scan to estimate the skull acoustic properties. The simulation is then rotated and interpolated to a global volume that includes the focal region and saved. After repeating for every element, the combined field can be rapidly obtained with different phase/magnitude values for the elements.

Before treatment, a CT scan of the head was obtained with a bone reconstruction kernel and a slice thickness of 1 mm or less. On the day of treatment, anatomic MRI scans were acquired in three orientations using a 3D FIESTA sequence (TR/TE: 5.1/2.4 ms; flip angle: 55°; receiver bandwidth: ±27.3 kHz; field of view: 22 cm; slice thickness 2 mm; matrix (typical) 224×288×28). The CT and MRI were registered to each other using the software of the ExAblate system. Phase aberrations were calculated using a proprietary method by the manufacturer. The amplitudes of each phased array element were set by the device software to normalize the acoustic exposure over the head to avoid hotspots. The transformation between the CT and MRI space, the target locations in MRI space, and the magnitude and phase of each transducer element were saved for each treatment. This information was used to simulate the acoustic field as described below.

During each sonication, MRTI was obtained in a time-series in a single, operator-defined plane using a fast spoiled gradient echo sequence. The orientation of the imaging plane and frequency/phase encoding directions varied between each sonication. Two acquisitions were obtained before the start of each sonication; the first was ignored due to artifacts. Acquisition continued over the course of the sonication; five additional acquisitions were obtained to map cooling. In the first 17 patients, MRTI parameters were: TR/TE: 27.8/12.9 ms; flip angle: 30°; receiver bandwidth ±5.7 kHz; field of view: 28 cm; slice thickness: 3 mm; matrix: 256×128. In the others, a multi-echo readout was performed (TR: 28.0 ms; TE: 3.1/7.8/12.5/17.2/21.9 ms; flip angle: 30°; receiver bandwidth: ±35.71 kHz; field of view: 28 cm; slice thickness: 3 mm; matrix: 256×128). The higher readout bandwidth of the multi-echo sequence reduced spatial distortions in the frequency-encoding direction with minimal loss of SNR [22].

The MRI reconstructed magnitude, real and imaginary images that were converted to phase-difference images. Temperature changes were estimated from the phase-differences [23] using a temperature dependence of -0.00909 ppm/°C [24]. This value was selected because it is what is used in the device software. The raw images were stored in DICOM format and independently analyzed offline using MATLAB software developed in-house. Artifacts resulting from changes in magnetic field (presumably due to patient motion outside the brain) were removed using a procedure outlined elsewhere where non-heated regions in each phase-difference images were fit to a smooth 2D surface [25]. The surface was extrapolated into the heated region and subtracted off. The phase difference maps obtained with the multi-echo sequence were combined via a weighted average based on the expected signal-to-noise ratio of the MRTI [26], with the weight for the nth echo given by:

$$W_n = \left(TE_n \cdot e^{-TE_n/T2^*}\right)^2 \qquad (1)$$

We used 30 ms for the $T2^*$ relaxation time for brain.

Table 1. Acoustic and thermal parameters used in the numerical model

| Parameter | Water | Brain | Skull |
|---|---|---|---|
| Density (mg/kg³)[a] | 1000 | 1030 | from CT |
| Sound speed (m/s)[a] | 1500 | 1560 | from density |
| Attenuation at 660 kHz (Np/m)[b] | 0 | 4.36 | from density |
| Thermal conductivity [W/(m·°K)] [a] | - | 0.51 | - |
| Specific heat [J/(kg·°K)] [a] | - | 3640 | - |
| Perfusion coefficient (l/s) [a] | - | 8.33E-03 | - |

[a]Ref. [28]; [b]Ref. [29]

## B. Simulations

Numerical modelling of the acoustic field was performed using k-Wave, an open-source MATLAB toolbox [21]. For every patient, the pressure amplitude was simulated individually for each element of the array. The element was assumed to be a 1 cm diameter flat circular piston centered on the location provided by the manufacturer. The dimensions of the elemental simulations were 44×44×492 elements with a spacing of 0.325 mm, which resulted in a simulation space of 14.3×14.3×159.7 mm³ per element. We used a perfectly matched layer with a size of 10 grid points and an attenuation of 2 Np per grid point. After the completion of the individual simulations, we transformed the resulting pressure field from the element space to a global space in the *xyz* frame of the transducer with dimensions of 14.3×14.3×23 mm³ via cubic interpolation. For a given sonication, the 44×44×71×993 matrix of elemental simulations was loaded, and the phase corrections supplied by the manufacturer were applied along with phase values needed to electronically steer the focus to the target location. Phase corrections for "ideal" focusing were found by simply subtracting the phase at the target location from each element's simulated pressure.

Next, we used the bioheat equation to estimate the heating [20]. We averaged the heating to match the voxel dimensions and temporal resolution of the MRTI with the imaging orientation used during treatment.

The parameters used in the numerical modelling of the acoustic pressure and the subsequent temperature rise are listed in Table 1. To estimate the sound speed and attenuation of the skull, we first used the density estimated from the CT scans and the empirical relationships presented by Pichardo et al. [15] interpolated to 660 kHz. We assumed a linear relationship between Hounsfield units and density, and -1000 and 57 Hounsfield units for air and soft tissue, respectively.

Overall, we considered 72 patient treatments. The first 32 patients were examined in detail and were used to optimize the relationships between skull density and the acoustic attenuation and sound speed. Thirty of these

Table 2. Patient information, treatment parameters, and CT scan settings

| Patient | | | Treatment | | | | | | CT Scan | | | |
|---|---|---|---|---|---|---|---|---|---|---|---|---|
| N | Age | SDR | # son. | Power (W) | Duration (s) | Acoustic energy (kJ) | | | Vendor | Kernel | keV | Voxel (mm) |
| | | | | | | Max. | Total | To reach 55°C | | | | |
| 1 | 58M | 0.61 | 17 | 139-797 | 10-21 | 12.9 | 107.6 | 8.0 | SIEMENS | H60s | 120 | 0.47×0.47×1.00 |
| 2 | 73F | 0.51 | 17 | 139-694 | 8-27 | 16.9 | 112.1 | 14.3 | SIEMENS | H60s | 120 | 0.46×0.46×1.00 |
| 3 | 58M | 0.41 | 15 | 185-937 | 10-16 | 13.9 | 99.7 | 7.7 | TOSHIBA | FC30 | 120 | 0.43×0.43×1.00 |
| 4 | 65F | 0.47 | 16 | 185-937 | 10-28 | 22.4 | 160.7 | 8.8 | SIEMENS | H60s | 120 | 0.52×0.52×1.00 |
| 5 | 78M | 0.41 | 17 | 186-934 | 10-20 | 18.6 | 125.2 | 9.5 | SIEMENS | H60s | 120 | 0.48×0.48×1.00 |
| 6 | 72F | 0.48 | 15 | 187-932 | 8-10 | 9.2 | 66.8 | 4.2 | SIEMENS | H60s | 120 | 0.43×0.43×1.00 |
| 7 | 65M | 0.49 | 13 | 186-1095 | 10-10 | 10.8 | 65.4 | 5.3 | SIEMENS | H60s | 120 | 0.38×0.38×1.00 |
| 8 | 69M | 0.40 | 25 | 188-1197 | 6-44 | 44.3 | 418.0 | 34.6 | SIEMENS | H60s | 120 | 0.56×0.56×1.00 |
| 9 | 68M | 0.46 | 15 | 187-840 | 10-24 | 18.8 | 114.4 | 10.0 | TOSHIBA | FC30 | 120 | 0.50×0.50×1.00 |
| 10* | 73M | 0.46 | 24 | 187-1198 | 10-23 | 27.4 | 193.1 | 25.4 | TOSHIBA | FC30 | 120 | 0.47×0.47×1.00 |
| 11* | 48M | 0.50 | 13 | 185-1116 | 10-13 | 14.4 | 90.1 | 12.2 | SIEMENS | H60s | 120 | 0.48×0.48×1.00 |
| 12 | 77F | 0.44 | 14 | 189-473 | 7-24 | 9.5 | 73.0 | 5.8 | SIEMENS | H60s | 120 | 0.43×0.43×1.00 |
| 13 | 83F | 0.42 | 12 | 185-1180 | 10-13 | 15.2 | 77.9 | 8.3 | SIEMENS | H60s | 120 | 0.43×0.43×1.00 |
| 14 | 80M | 0.38 | 12 | 235-1292 | 10-16 | 21.1 | 121.8 | 12.7 | SIEMENS | H60s | 120 | 0.52×0.52×1.00 |
| 15 | 70M | 0.50 | 9 | 230-939 | 10-13 | 10.8 | 51.7 | 5.4 | SIEMENS | H60s | 120 | 0.50×0.50×1.00 |
| 16 | 80F | 0.38 | 11 | 232-1030 | 10-24 | 24.3 | 101.2 | 15.3 | SIEMENS | H60s | 120 | 0.45×0.45×1.00 |
| 17 | 91M | 0.51 | 10 | 230-1026 | 10-13 | 13.2 | 48.1 | 7.4 | SIEMENS | H60s | 120 | 0.46×0.46×1.00 |
| 18 | 71M | 0.55 | 13 | 190-853 | 10-16 | 13.0 | 67.6 | 5.4 | SIEMENS | H60s | 120 | 0.47×0.47×1.00 |
| 19 | 75M | 0.69 | 12 | 142-837 | 8-13 | 9.5 | 51.6 | 4.4 | TOSHIBA | FC30 | 120 | 0.47×0.47×1.00 |
| 20 | 78M | 0.45 | 20 | 238-807 | 9-32 | 21.4 | 172.6 | 13.3 | SIEMENS | H60s | 120 | 0.50×0.50×1.00 |
| 21 | 79F | 0.57 | 12 | 189-662 | 10-21 | 9.5 | 56.6 | 3.7 | SIEMENS | H60s | 100 | 0.39×0.39×1.00 |
| 22 | 71M | 0.44 | 11 | 190-1123 | 9-24 | 19.0 | 85.4 | 10.8 | SIEMENS | H70h | 120 | 0.49×0.49×0.50 |
| 23 | 69F | 0.46 | 13 | 191-1148 | 10-33 | 36.1 | 181.6 | 23.3 | SIEMENS | H37s | 120 | 0.48×0.48×1.00 |
| 24 | 84M | 0.55 | 10 | 237-792 | 11-17 | 9.1 | 54.3 | 4.6 | GE | BONE+ | 120 | 0.49×0.49×1.00 |
| 25 | 77M | 0.59 | 10 | 189-707 | 10-20 | 13.4 | 61.0 | 6.4 | GE | BONE+ | 140 | 0.53×0.53×0.63 |
| 26 | 86F | 0.41 | 10 | 189-847 | 10-19 | 15.3 | 87.4 | 6.4 | SIEMENS | H60s | 120 | 0.41×0.41×1.00 |
| 27 | 72M | 0.61 | 12 | 189-903 | 11-15 | 12.6 | 77.8 | 5.7 | GE | BONE+ | 120 | 0.54×0.54×0.63 |
| 28 | 84M | 0.40 | 13 | 191-1302 | 11-36 | 31.5 | 186.8 | 14.3 | SIEMENS | H60s | 120 | 0.45×0.45×1.00 |
| 29 | 72F | 0.45 | 11 | 89-953 | 12-26 | 23.6 | 101.3 | 7.3 | TOSHIBA | FC30 | 120 | 0.47×0.47×1.00 |
| 30 | 81M | 0.48 | 13 | 188-848 | 12-33 | 22.2 | 160.9 | 13.9 | GE | BONE+ | 140 | 0.49×0.49×1.00 |
| 31 | 79M | 0.60 | 12 | 183-1054 | 6-14 | 13.2 | 61.3 | 5.4 | SIEMENS | H60s | 120 | 0.52×0.52×1.00 |
| 32 | 83M | 0.66 | 12 | 93-566 | 12-20 | 6.9 | 46.9 | 4.4 | GE | BONE+ | 120 | 0.49×0.49×0.63 |

*Pallidotomy

patients were treated in the ventral intermediate (VIM) nucleus of the thalamus for essential tremor; two were treated in the globus pallidus for Parkinson's disease. The number of sonications per treatment and the acoustic parameters used (power, duration) varied between patients. Details for each patient, their treatments, and the CT scans are listed in Table 2. A total of 447 sonications were performed in the treatments in patients 1-32; 16 sonications were excluded due to artifacts in the focal region in MRTI. After completing the analyses of the first 32 patients, we validated our results by simulating 40 additional essential tremor patients (Table S1).

The simulations were performed in MATLAB using the O2 High Performance Compute Cluster, supported by the Research Computing Group at Harvard Medical School. We stored the complex steady-state pressure amplitude after each simulation. The simulations in the first 32 patients were run five times: using the density/attenuation relationship described in Pichardo et al. [15], without attenuation, using the optimized density/attenuation and finally with the two optimized density/sound speed relationships. The optimization procedure is described below. We simulated the additional 40 patients with the two optimized density/sound speed relationships and without attenuation.

## C. Optimizing skull attenuation

We investigated the feasibility of systematically iterating the elemental simulations to find a relationship between attenuation and skull density that resulted in heating that better matched the MRTI for the first 32 patients. It was impractical to repeatedly run the full numerical model, so we used a simplified attenuation model that we applied to simulations performed without attenuation. This simplified model assumed that the primary attenuation effects occur along the z-direction:

$$P_k(x_i, y_i, z_i) \approx P_{k0}(x_i, y_i, z_i)$$
$$\times \exp\left[-\sum_{j=0}^{i} \alpha(x_i, y_i, z_j)\Delta z\right], \quad (2)$$

where $P_{k0}(x_i, y_i, z_i)$ is the pressure distribution simulated without attenuation for element $k$, $\alpha(x_i, y_i, z_i)$ is the attenuation, and $\Delta z$ is the grid size.

We assumed the relationship between attenuation and skull density $\rho_{sk}(x_i, y_i, z_i)$ can be approximated by a series of polynomials:

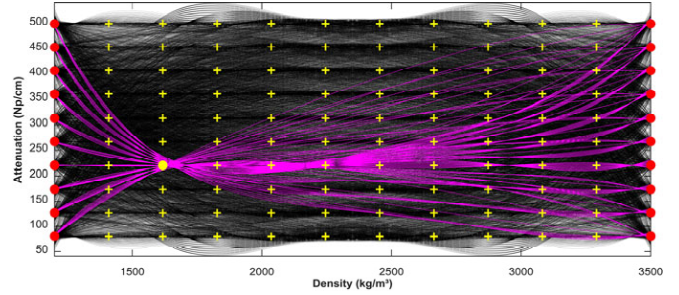

FIG. 2. Optimizing the relationships between skull density and attenuation. We examined 10,000 relationships per sonication (black curves). To generate the curves, we used two nodes with densities of 1200 and 3500 kg/m³ (red circles) and a third indicated by one of the yellow crosses. These nodes were fit to a polynomial. We generated a set of 100 curves for each yellow cross; an example set is shown in magenta.

$$\alpha_{sk}(x_i, y_i, z_i) \approx \sum_m A_m \rho_{sk}(x_i, y_i, z_i)^m, \quad (3)$$

resulting in:

$$P_k(x_i, y_i, z_i) \approx P_{k0}(x_i, y_i, z_i) \exp[-N_b(x_i, y_i, z_i)\alpha_b\Delta z]$$
$$\times \exp\left[-\sum_{j=0}^{i}\sum_m A_m \rho_{sk}(x_i, y_i, z_j)^m \Delta z\right] \quad (4)$$

where $N_b(x_i, y_i, z_i)$ is the number of soft tissue voxels between the transducer and $z_i$. After exchanging the order of summation this becomes:

$$P_k(x_i, y_i, z_i) \approx P_{k0}(x_i, y_i, z_i) \exp[-N_b(x_i, y_i, z_i)\alpha_b\Delta z]$$
$$\times \exp\left[-\sum_m A_m \Delta z \sum_{j=0}^{i} \rho_{sk}(x_i, y_i, z_j)^m\right] \quad (5)$$

The values for $N_b(x_i, y_i, z_i)$ and the cumulative summations of the skull density $(\sum_{j=0}^{i} \rho_{sk}(x_i, y_i, z_j)^m)$ for orders $m$=0-4 were calculated for each transducer element, interpolated to the array space, and saved. These data, along with the individual pressure estimates for each element could then be loaded into memory. The total three-dimensional pressure amplitude for all elements and the resulting temperature rise could then be estimated for different values of $A_m$. We only included the 23×23×31 spatial simulation points centered on the focus location of the simulation to save time.

We attempted to find the $A_m$ coefficients that minimized the difference between the simulated and measured temperature rise for all patients. It was not practical to load the simulated pressure fields and skull density summations for the 32 patients into computer memory simultaneously. We thus systematically explored different coefficients for each patient separately and in parallel. We defined a grid of nodes at 12 values of density between 1200 and 3500 kg/m³ and 10 values of attenuation between 80 and 500 dB/cm (FIG. 2). We then selected a set of three nodes, which were each then fit to a $4^{th}$ order polynomial to determine the $A_m$ coefficients. Two of these nodes had densities of 1200 and 3500 kg/m³ (100 possible sets), and the third was one of the 100 nodes between these two. Thus, for each patient we examined 10,000 sets of $A_m$ coefficients.

For each set of coefficients, we used equation (5) to estimate the attenuated pressure field for the individual elements. We then set the phase and magnitude for each sonication and estimated the focal temperature rise. After compiling these temperature estimates, we found the coefficients that minimized the mean squared error between the simulated and measured temperature rise. Finally, we repeated this full simulations with the optimized attenuation/density relationship.

As shown in the results below, in many patients, plots of the difference between measured and simulated heating as a function of previously applied energy was relatively constant for the initial sonications, after which it began to deviate. We interpreted this deviation as an irreversible change in skull properties. Thus, when we found the optimized $A_m$ coefficients, we only included sonications that occurred before this deviation.

### D. Optimizing skull sound speed

The deviations between simulations and MRTI in plots of heating as a function of acoustic energy suggest that the skull acoustic properties might be changed during treatment. Assuming *ad hoc* that this change was primarily in sound speed, we investigated whether we could find optimized relationships between sound speed and skull density that could be used before and after this change occurred. As described above for attenuation, we created a simplified model that allowed us to rapidly explore different density/sound speed relationships using previously-obtained simulations. We assumed that the relationship between the inverse of the sound speed and the skull density can be approximated by a series of polynomials:

$$\frac{1}{c_{sk}(x_i, y_i, z_i)} = \sum_m B_m \rho_{sk}(x_i, y_i, z_i)^m \quad (6)$$

To estimate the effect of changing the skull sound speed on the transmitted pressure magnitude, we assumed the biggest loss occurred due to the initial reflections at the interfaces between soft tissue, the inner and outer tables, and the diploe (i.e. we ignored multiple reflections and other reflections within the skull). The transmission coefficient was calculated using the following relationship [12]:

$$\begin{aligned} T_k(x_i, y_i) &\approx \left( \frac{2Z_{OT}\cos\theta_i\cos\theta_t}{Z_{OT}\cos\theta_i + Z_T\cos\theta_t} \right) \\ &\times \left( \frac{2Z_{DP}\cos\theta_i'\cos\theta'_t}{Z_{DP}\cos\theta_i' + Z_{OT}\cos\theta_t'} \right) \\ &\times \left( \frac{2Z_{IT}\cos\theta_i''\cos\theta''_t}{Z_{IT}\cos\theta_i'' + Z_{DP}\cos\theta_t''} \right) \\ &\times \left( \frac{2Z_T\cos\theta_i'''\cos\theta'''_t}{Z_T\cos\theta_i''' + Z_{IT}\cos\theta_t'''} \right) \end{aligned} \quad (7)$$

where the acoustic impedance of the outer and inner tables, the diploe, and soft tissue, respectively, are given by:

$$\begin{aligned} Z_{OT} &= \rho_{OT}(x_i, y_i)c_{OT}(x_i, y_i) \\ Z_{IT} &= \rho_{IT}(x_i, y_i)c_{IT}(x_i, y_i) \\ Z_{DP} &= \rho_{DP}(x_i, y_i)c_{DP}(x_i, y_i) \\ Z_T &= \rho_T c_T \end{aligned} \quad (8)$$

We used brain values for the soft tissue density ($\rho_T$) and sound speed ($c_T$), and density values measured at the outer and inner tables of the skull and the diploe for $\rho_{OT}(x_i, y_i)$, $\rho_{IT}(x_i, y_i)$, and $\rho_{DP}(x_i, y_i)$, as described below. The incident and transmitted angles with respect to the normal vectors at the outer and inner tables and the diploe ($\theta_i, \theta_t, \theta_i', \theta_t'$, respectively) were found via Snell's law:

$$\begin{aligned} \frac{\sin\theta_t}{\sin\gamma_O} &= \frac{\sin\theta_t}{\sin\theta_i} = \frac{c_{OT}(x_i, y_i)}{c_T} \\ \frac{\sin(\theta_t')}{\sin(\gamma_O - \gamma_{DP} + \theta_t)} &= \frac{c_{DP}(x_i, y_i)}{c_{OT}(x_i, y_i)} \\ \frac{\sin(\theta_t')}{\sin(\theta_t')} &= \frac{c_{IT}(x_i, y_i)}{c_{DP}(x_i, y_i)} \\ \frac{\sin(\theta_t''')}{\sin(\gamma_I - \gamma_{DP} + \theta_i'')} &= \frac{c_T}{c_{IT}(x_i, y_i)} \end{aligned} \quad (9)$$

where $\gamma_O$, $\gamma_{DP}$ and $\gamma_I$ were the incidence angles of the outer skull surface, the center of the diploe, and the inner skull surface measured with respect to the z-axis. A diagram showing how the incidence angles were defined for this procedure is shown in FIG. S1. We measured the density and angles of incidence for the inner and outer tables with respect to the z-axis, for each $(x_i, y_i)$ coordinate as described below.

To estimate phase shifts resulting from changing the skull sound speed, we multiplied the delays expected the along the z-direction by $2\pi$ times the FUS frequency ($f$):

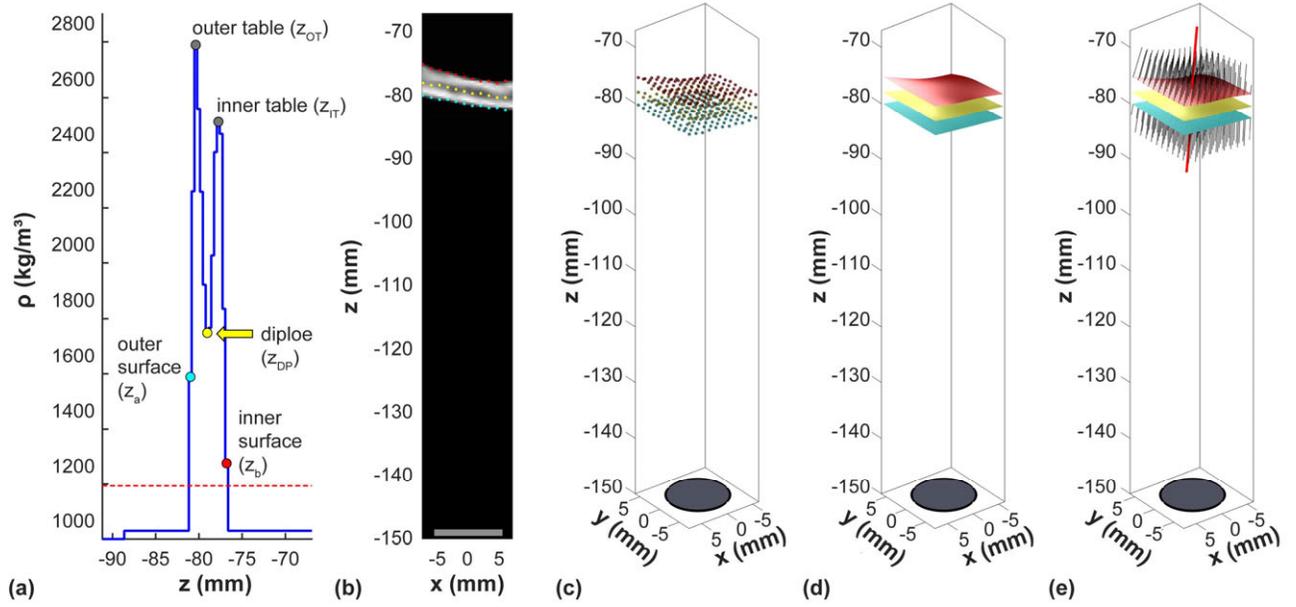

FIG. 3. Feature extraction from the CT scans for each phased array element. (a) For every x and y coordinates of the elemental simulation, threshold analysis of the density along the z-direction was used to find the first and last voxel of the inner and outer skull surface. Working from the outer surface, the first peak encountered was used to identify the outer table. The same procedure was used working from the inner surface to find the inner table. The minimum density between the inner and outer table was used to identify the diploe. The coordinates of the inner and outer surface and at the diploe (b, c) were fit to smooth surfaces (d). The normal vectors for these surfaces were found for each x and y coordinate (e); the individual normal vectors (black) and the mean vector (red lines) are shown. Note that for clarity, (b), (c), and (e) only plot every fourth x and y coordinate.

$$\Delta\varphi_k(x_i, y_i) \approx 2\pi f \sum_{j=0}^{i} \frac{\Delta z}{c(x_i, y_i, z_j)} \qquad (10)$$

After combining with eq. 6 and changing the order of summation, this becomes:

$$\Delta\varphi_k(x_i, y_i) \approx 2\pi f \Delta z \sum_m B_m \sum_{j=0}^{i} \rho_{sk}(x_i, y_i, z_j)^m \qquad (11)$$

The cumulative summations of the skull density ($\sum_{j=0}^{i} \rho_{skull}(x_i, y_i, z_j)^m$) saved for the attenuation estimation were used.

We applied the magnitude and phase changes to simulations performed with the optimized attenuation. We used eq. 7 and 11 to remove the effects of the density/sound speed relationship used in those simulations before testing new ones.

As was done with the attenuation optimization, we estimated the temperature rise over a predefined set of density/sound speed curves for every sonication. We defined a grid of curves covering densities between 1200 and 3500 kg/m³ and sound speeds between 1500 and 4000 m/s. We then selected a set of three nodes, and we fit the inverse of the sound speed to a 4th order polynomial to determine the $B_m$ coefficients. Two of these nodes had densities of 1200 and 3500 kg/m³ (100 possible sets), and the third was one of the 100 nodes between these two. Thus, for each patient we examined 10,000 sets of $B_m$ coefficients. The relationship that minimized the error between the simulations and measurements was found for the sonications that deviated from the model after attenuation optimization. We then repeated the full simulations with the optimized attenuation and sound speed relationships.

### E. Data Analysis

All analysis was performed in MATLAB. We compared the peak temperature rise at the focus measured with MRTI to that estimated with the simulations using linear regression. To compare "treatment efficiency", we estimated the energy needed in each patient to achieve a focal temperature of 55°C. We noted that plots of measured focal heating as a function of acoustic energy generally followed a logarithmic curve. Thus, we used nonlinear least squares regression (using the function "nlinfit" in MATLAB) to estimate this energy. To further test our ability to predict

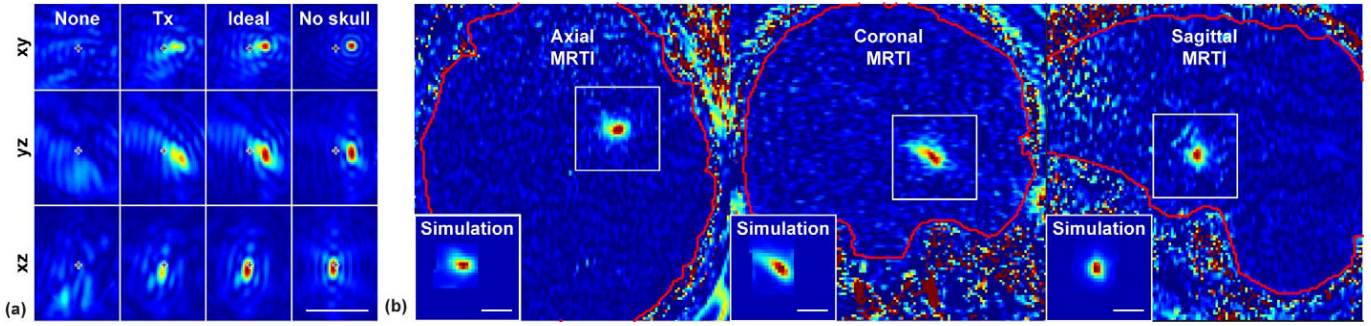

FIG. 4. Example simulated pressure distributions (a) and measured and predicted temperature maps (b) from a pallidotomy treatment (patient 11). In (a), three orientations with respect to the transducer are shown for one sonication with no aberration correction, the correction used during treatment (Tx), and ideal correction; the simulated field without a skull (but with magnitude shading and beam steering) is also shown. The location of the geometric focus is indicated. For this sonication, the focus was electronically steered away from the geometric focus to x = 3 mm and y = 1 mm. In (b), MRTI for three sonications are shown; the insets are the simulated heating with the aberration correction used during treatment for these examples. Scale bars: 1 cm

the "treatment efficacy", we compared the measured and predicted slopes of plots of the logarithm of the acoustic energy and the focal temperature rise.

We also compared the shape and orientation of the heating. The simulated and measured temperature maps at the time of peak temperature rise were fit using the Levenberg-Marquardt method via the MATLAB function "fit.m" to a two-dimensional Gaussian distribution:

$$
\begin{aligned}
&\Delta T(x, y) \\
&= A + B \times \exp\left[-\frac{(x - x_0)\cos\theta + (y - y_0)\sin\theta}{\sigma_x}\right]^2 \\
&\times \exp\left[-\frac{(x - x_0)\sin\theta + (y - y_0)\cos\theta}{\sigma_y}\right]^2
\end{aligned}
\tag{12}
$$

We used linear regression to compare the resulting estimates of the dimensions ($\sigma_x, \sigma_y$) and obliquity ($\theta$). For the obliquity, not all the temperature maps had a well-defined angle. Thus, values for $\theta$ that had 95% confidence intervals greater than ±30° were excluded. In each patient, we selected representative sonications that were largely free of artifacts in the MRTI. We selected examples from each imaging orientation and frequency encoding direction for each patient. Not all orientations and encoding directions were available in every patient. Overall, we examined 110 sonications when comparing the focal heating shape.

We examined whether different factors gleaned from the skull CT could predict the acoustic energy needed to achieve an effective thermal exposure at the focus (i.e. energy required to heat to 55°C). These factors were determined with the individual volumes used in the element-wise simulations (FIG. 3). Working along the direction of propagation, we identified the first and last skull voxel, the voxels at the center of the inner and outer tables, and the voxel that had the minimum density between the outer and inner table for each of the 44×44 coordinates in the element-wise simulation. The coordinates of the first and last skull voxel and the diploe were fit to smooth surfaces. Normal vectors to this surface were calculated to obtain the angles of incidence of the skull and the direction of ultrasound propagation. We used these data to calculate different skull-derived factors (see Supplemental Methods for more details). We then examined the relationship between each factor and the energy needed to reach a peak focal temperature of 55°C. The ability of different metrics to predict this energy were evaluated using linear regression with the "fitlm.m" function in MATLAB. Since there were obvious outliers, we used the robust fitting option which uses iteratively re-weighted least squares with a bi-square weighting function.

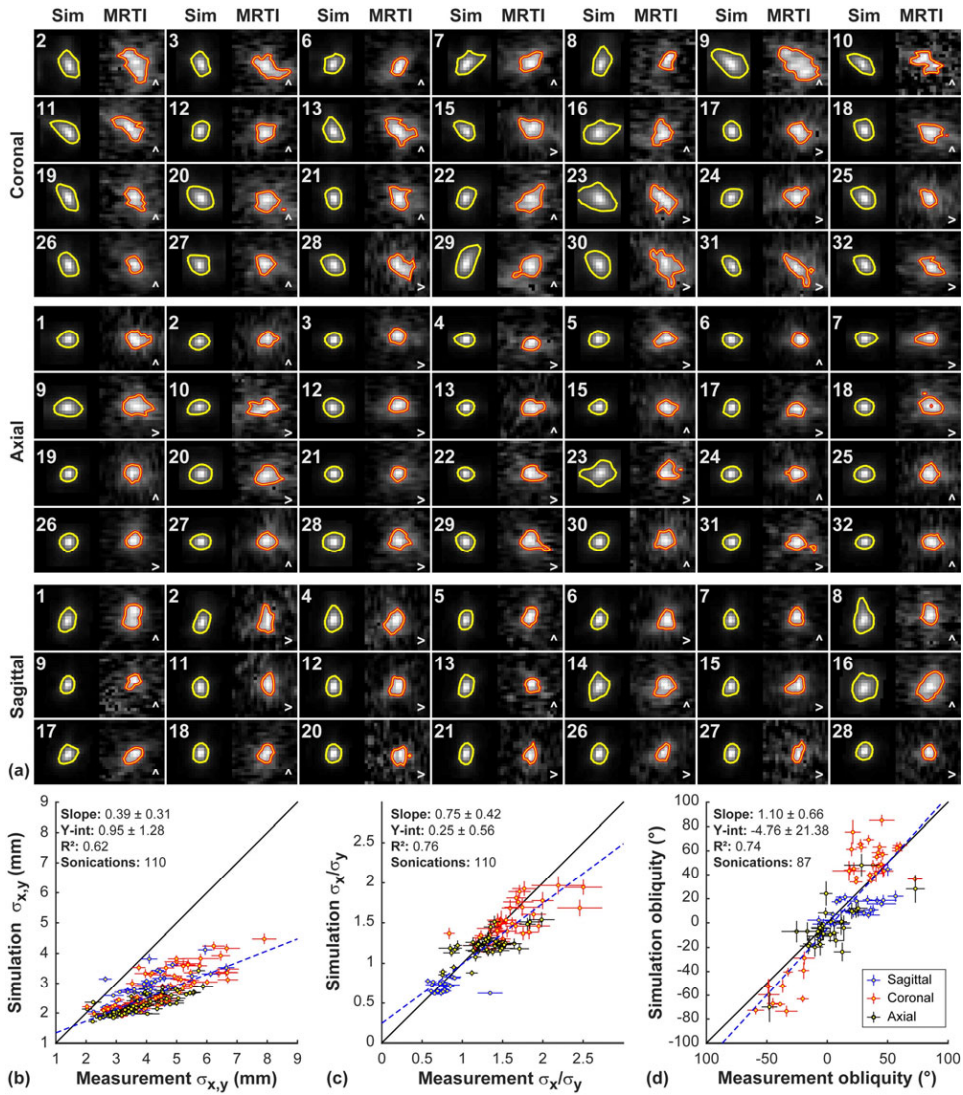

FIG. 5 Comparing predicted and measured focal shape, size, and orientation. (a) Simulated temperature maps (Sim) and corresponding MRTI acquired during TcMRgFUS in 32 patients. One example from each imaging orientation was selected per patient; not every orientation was used in every patient. The shape, relative size, and obliquity of the simulations matched the measurements reasonably well in most cases. However, the measured focal heating was more diffuse than the simulations predicted. The orange lines indicate the 50% contours for the MRTI; yellow lines indicate 25% contours for the simulations. The field of view of each region shown is 2.3 cm. The patient number and the frequency-encoding direction in the MRTI are indicated. (b-d) Simulated and measured heating dimensions and obliquity. The simulated and measured MRTI were both fit to a two-dimensional Gaussian distribution. (b) A good correlation was observed between the dimensions ($\sigma_x$, $\sigma_y$) of these fits, but the dimensions in MRTI were higher. (c) A good correlation was observed in the ratios of widths. (d) In many cases this fit detected tilted heating distributions. The simulations detected a tilt in the same direction. (error bars: 95% CI of the fits; dotted line: linear regression; solid line: unity)

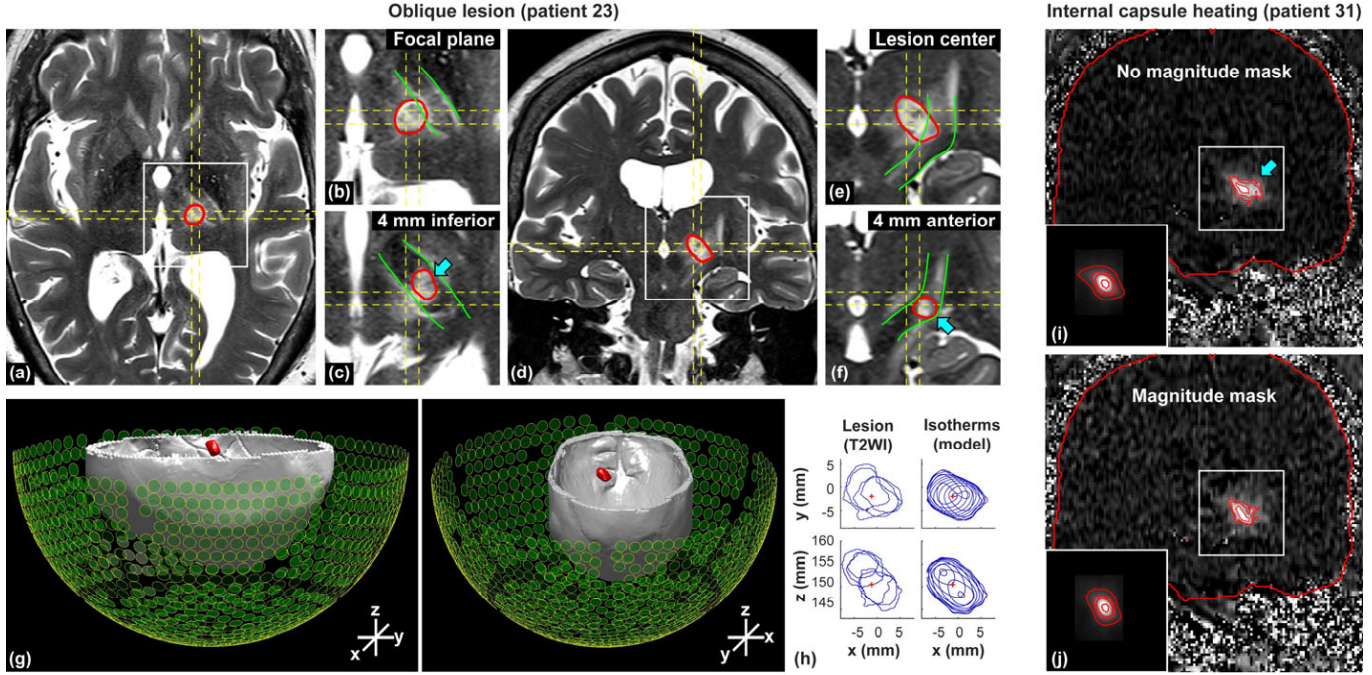

FIG. 6. Risks of an oblique focus. (a-h) Tilted lesion in an essential tremor patient. (a-f) T2-weighted MRI 24h after FUS (a-c: axial; d-f: coronal). The MRTI imaging planes are indicated by dashed lines, the lesion boundary is segmented in red, and the internal capsule is indicated in green. Coronal imaging through the lesion center (d, e) revealed a left/right tilt. Axial imaging inferior to the lesion center (c) and coronal images anterior to it (d) show that the it was also tilted in the superior/inferior direction and included a portion of the internal capsule. This portion of the lesion (arrows) was not included in MRTI in any orientation. (g) 3D rendering of isotherms generated by simulation (red) along with calvaria (from CT) and the transducer array. The predicted heating is tilted both in the left/right and superior/inferior directions. (h) Stacks of contours from the lesion segmented in MRI and the simulated isotherms from two different viewpoints. Both the lesions and the isotherms had similar obliquity. (i-j) Heating in the internal capsule in a different patient. MRTI and simulated heating (insets) during two consecutive sonications with and without a magnitude mask applied to reduce left/right obliquity of the focus. This reduction is evident in both the MRTI and the simulations. However, the simulations did not predict the lateral heat spread into the internal capsule (arrow).

## III. RESULTS

### A. Comparing shape, size, and obliquity of simulated and measured temperature maps

FIG. 4 shows examples of the simulated pressure and the corresponding measured and simulated focal heating during TcMRgFUS, here in a pallidotomy procedure. This patient was selected due to the large left-right tilt in the focal heating. Pressure distributions, displayed in the *xyz* planes of the transducer, are shown with the phase corrections used during treatment, ideal corrections where the pressure for each element have equal phase at the target, and no correction. The simulated temperature maps are shown after spatial averaging to match the planes of the MRTI. The relative sizes and the obliquity of the focal heating are similar to the measurements.

This general agreement was evident overall. FIG. 5a shows example simulated temperature maps and corresponding MRTI for patients 1-32 in different imaging orientations. The relative size and obliquity of the focal heating were similar in most patients. However, the heating was more diffuse in the MRTI than was predicted by the simulations; FIG. 5a shows 50% isotherms in the MRTI and 25% isotherms in the simulations.

To characterize the ability of the simulations to predict the shape of the heating measured with MRTI, we estimated the dimensions and obliquity by fitting the focal heating in the MRTI and the simulated temperature maps to a two-dimensional Gaussian distribution (FIG. 5b-d). Both the dimensions and tilt angles measured in the MRTI correlated strongly with those predicted by the simulations ($R^2$: 0.62 and 0.74, respectively; $P < 0.001$). However, the dimensions of the simulations were smaller than the measurements in most cases. The simulations predicted a tilt in the same direction as the MRTI, but the absolute angle was less in the simulations, particularly in the left/right direction in coronal MRTI. A good correlation (R²: 0.76) was observed for the ratio of the dimensions ($\sigma_x/\sigma_y$) of these fits.

Left/right obliquity and elongation in the superior/inferior direction poses a risk during thalamotomy of damaging the internal capsule, which is located laterally and inferior to the location of the thalamic target. An extreme example of such a case is shown in FIG. 6a-h. In this patient, there was obliquity in both the left/right and superior/inferior directions, leading to heating in the internal capsule that could not have been detected in the single-plane MRTI in any orientation (FIG. 6c,f). This double obliquity was predicted by the simulations (FIG. 6g-h). Another example of this risk is shown in FIG. 6i-j. The device manufacturer added an option to change the magnitude distribution of the phased array to reduce the left/right obliquity. FIG. 6i-j shows MRTI acquired with and without this magnitude mask. While a corresponding reduction in obliquity is also evident in the simulations, the shape of the predicted heating is less diffuse and does not include the lateral heat spread into the internal capsule (arrow in FIG. 6i).

## B. Predicting focal heating

The examples presented above demonstrate that the simulations generally predicted the relative size and shape of the focal regions that were observed in MRTI. We also compared measurements and simulations of the peak temperature at the focus for the 431 sonications delivered without MRTI artifacts in these 32 patients.

Our first simulations used the relationship between density and attenuation found by Pichardo et al. [15]. After comparing simulations and measurements (FIG. S2), we made two general observations. First, while in some cases good agreement was observed, we found that the simulated temperatures were generally less than the measurements, particularly when the exposure level was relatively low. Second, we noted that plots of acoustic energy vs. heating followed parallel trajectories during the initial sonications performed during the treatments. However, in many patients, these trajectories deviated as the acoustic energy increased, and the measured heating increased less than the simulations predicted as the energy increased.

One interpretation of these two observations is that the relationship between skull attenuation and density was incorrect, and that in some patients the acoustic properties of the skull changed after a certain level of

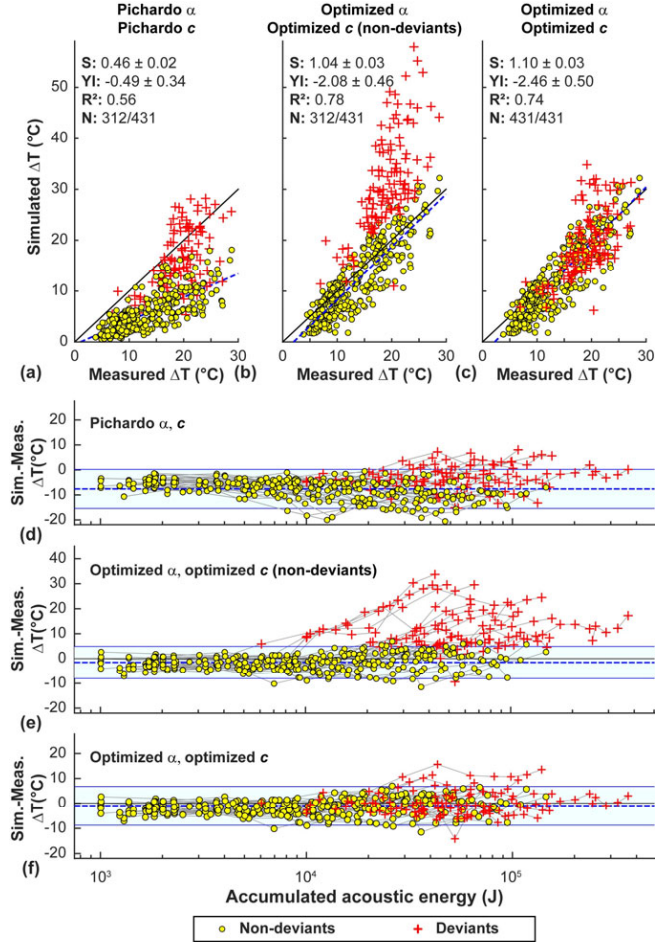

FIG. 7. Comparison of measured and simulated peak temperature rise for 431 sonications in 32 patients. (a-c) Comparison using the density/attenuation ($\alpha$) and density/sound speed ($c$) relationships found in Pichardo et al. [15], using optimized attenuation, and using optimized attenuation and sound speed. (d-f) Difference between the predicted and measured heating as a function of the accumulated acoustic energy. With the literature relationships (a, d), the simulations generally under-predicted the measured heating. Using the optimized attenuation relationship (b, e), the agreement was better. At low energies, the difference between the predictions and measurements was constant. However, in 15/32 patients, the simulations began to over-predict the focal heating as the energy increased. These deviant sonications (red crosses) were excluded from the attenuation optimization. Using the optimized density/sound speed relationship resulted in better agreement for the deviant sonications (c). The lines in (d-f) connect data obtained in each patient. (S, YI: slope, y-intercept of linear regression)

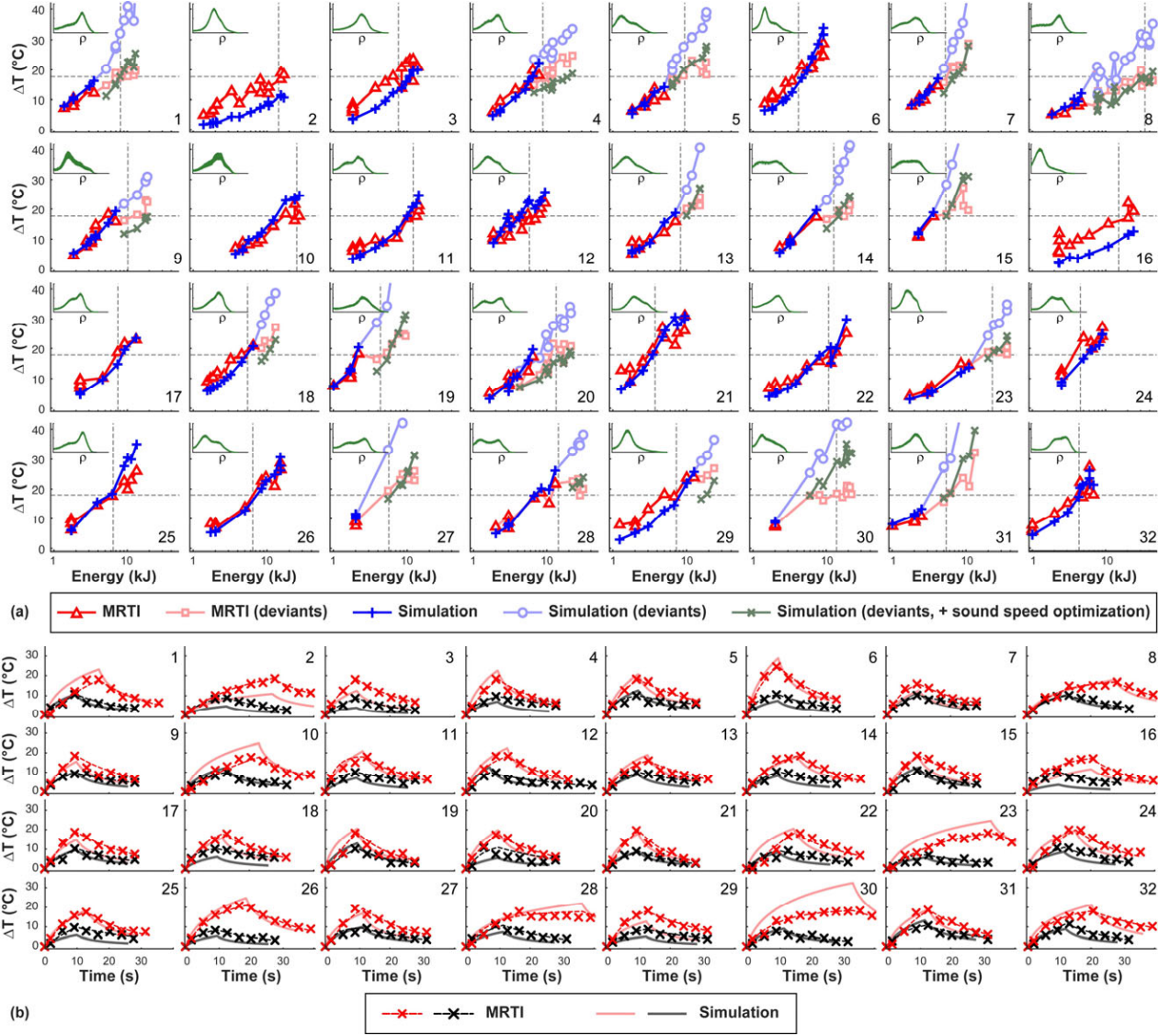

(a)

Legend: ■ MRTI ■ MRTI (deviants) ■ Simulation ■ Simulation (deviants) ■ Simulation (deviants, + sound speed optimization)

(b)

Legend: ✕✕ MRTI ✕✕ Simulation

FIG. 8. Comparing predicted and measured focal temperatures. (a) Measured and simulated focal heating as a function of the applied acoustic energy. For the sonications where the measurements and simulations deviated (red crosses in Figure 7), results from the two different density/sound speed optimizations are shown. The horizontal dotted lines indicate an absolute temperature of 55°C, a rough estimate for thermal necrosis; vertical lines indicate the energy needed to reach this threshold. The insets show histograms of skull density (x-axis: 1290-3500 kg/m³). (b) Measured and predicted temperature rise vs. time plots for two sonications selected from each patient. The two selected sonications had focal temperatures close to 47°C and 55°C during each treatment. The heating and subsequent cooling measured with MRTI were each fit to an exponential function. The apparent cooling rate for the simulations was higher than the measurements.

exposure, leading to defocusing. Thus, we evaluated the feasibility of optimizing the density/attenuation relationship to better match the MRTI for those sonications delivered before this change. FIG. 7a and d compares focal heating in MRTI to simulations using the attenuation/density relationship found by Pichardo et al. [15]. Results for the optimized attenuation are shown in FIG. 7b and e. The sonications indicative of a possible change in skull properties were evident after examining plots of the difference in heating between the simulations and the measurements as a function of the accumulated applied acoustic energy (FIG. 7d-f). Those deviant points, shown as red crosses, were selected automatically for each patient when two or more sonications exceeded a cut-off of two standard deviations above the mean difference (indicated by the light blue regions in FIG. 7d-f) for the optimized attenuation model. With the optimized model, 15/32 patients showed an obvious deviation in focal heating from the model. The mean previously-applied energy at which this deviation was $17.6 \pm 10.0$ kJ.

We next tested whether we could find a relationship between sound speed and density that predicted focal heating that better matched the measurements for the sonications both before and after the presumed change to the skull. The resulting estimated peak temperatures for the deviant sonications are shown with the red crosses in FIG. 7c and f. The agreement with the measurements was improved on average, but the heating in some patients was still not well predicted.

FIG. 8a shows plots of the measured and simulated temperature rise as a function of the acoustic energy for patients 1-32. With the optimized attenuation/density relationship, excluding the deviants, the trend in these plots was similar for measurements and simulations, and the simulations captured effects such as beam steering, changing sonication duration, and varying the plane used for the MRTI. These effects are evident in FIG. 8a where the simulated temperature/energy relationships are not smooth. The simulations grossly under-predicted the focal heating in a few patients (such as patients 2, 3, and 16). Furthermore, in most patients the measurements followed a logarithmic curve (appears linear in FIG. 8a, which has a logarithmic scale on the x-axis), while the simulations predicted a more linear relationship (appears curved in FIG. 8a). Patient 6 is an example of this behavior. FIG. 9 compares the slopes of the measured and predicted heating vs. acoustic energy curves in FIG. 8a. A good correlation between measurements and simulations was observed ($R^2$: 0.57), but the simulated efficiency was slightly higher in most patients.

Using the optimized density/sound speed relationship improved the prediction of the focal heating for the deviant sonications. The predicted focal heating with this optimized relationship is shown in FIG. 8a by the green "×" symbols. There were some patients (patients 29 and 30, for example) that were still not well-predicted by the simulations after this optimization.

As we gained experience with these treatments, we used larger steps in acoustic energy between sonications.

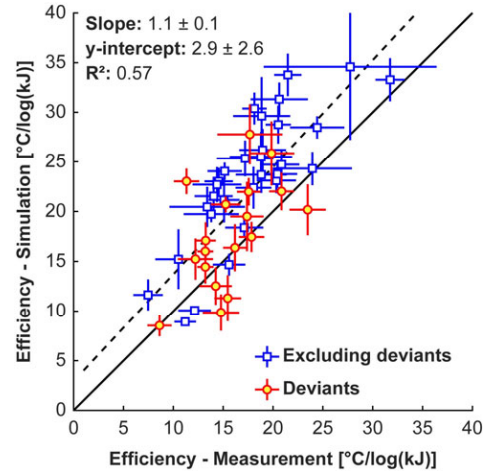

FIG. 9. Treatment efficiency. Comparisons of the measured and simulated slopes of the plots in FIG. 8 of focal heating as a function of the logarithm of the acoustic energy for 32 patients. The blue squares show the slopes excluding sonications where the measurement and simulated trajectories of the heating/acoustic energy plots deviated; red circles included those sonications. The error bars are the standard error of least-squares regression.

Thus, in two cases (patients 27 and 30), all sonications except the first few low-energy ones had heating less than the simulations predicted. In these two patients, there was a large difference between the simulation and the measurements even though the prior energy delivered was low.

## C. Focal heating vs. time

We compared plots of the temperature rise as a function of time for the measured and simulated MRTI (FIG. 8b). For each patient we selected two sonications, one where the heating was approximately 10°C, and another where the temperature rise was approximately 18°C (corresponding to an absolute temperature of 55°C). The heating rates of the simulations and measurements were similar. However, the decay of the heating after the sonication was faster in the simulations than the measurements. To characterize this, we fit the temperature decay to an exponential function and estimated the time required to reduce the heating by one half. The mean half-life was $14.0 \pm 3.7$ s in the measurements, significantly higher (P<0.001) than the $8.6 \pm 1.1$ s estimated in the simulations.

Table 3. Coefficients for eq. 3 and 6 that resulted in the best agreement between MRTI and the simulations

|  | $A_0$ | $A_1$ | $A_2$ | $A_3$ | $A_4$ |
|---|---|---|---|---|---|
| Before skull change: | 5.71E+03 | -9.02E+00 | 5.40E-03 | -1.41E-06 | 1.36E-10 |
|  | $B_0$ | $B_1$ | $B_2$ | $B_3$ | $B_4$ |
| Before skull change: | 3.68E-03 | -5.95E-06 | 4.13E-09 | -1.28E-12 | 1.48E-16 |
| After skull change: | 1.24E-03 | -7.63E-07 | 1.69E-10 | 5.31E-16 | -2.79E-18 |

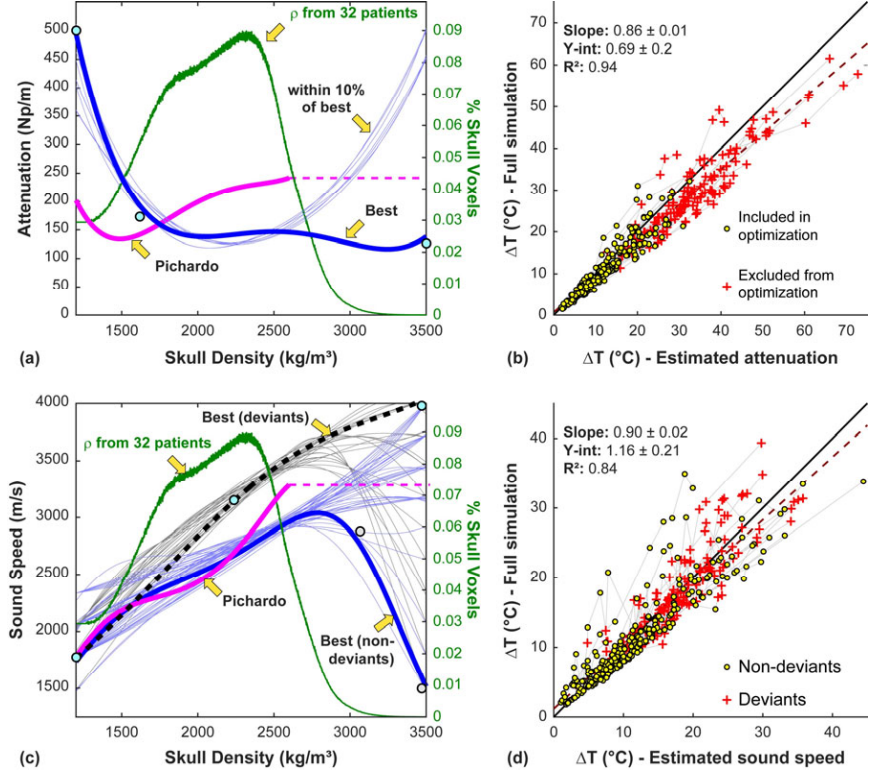

FIG. 10. Optimized skull density/attenuation and density/sound speed relationships. (a-b) For each of the 431 sonications, 10,000 different density/attenuation relationships were investigated using simulations performed without attenuation and a simplified attenuation model. The relationship that produced the best match to the MRTI and those that were within 10% of the best are shown. The attenuation model described by Pichardo et al. [15] is also shown. A histogram of the density of all the skulls in the study is shown in green. (b) Plot showing the heating predicted by the full simulations vs. that estimated using the simplified attenuation model in eq. 5. The simplified attenuation model slightly over-estimated the focal heating predicted by the full simulations. (dotted line: linear regression; solid line: unity). (c-d) Optimization of the density/sound speed relationship using a similar methodology. Here, the optimization was performed separately for the deviant and non-deviant sonications. (d) Focal heating estimated with the full simulation compared to the simplified model outlined in eq. 6-11. Good agreement on average was observed, although the simplified model deviated from the full model in some patients. The data points for the individual patients are connected in (b) and (d) by the pale grey lines.

### D. Optimizing relationships between skull density, attenuation, and sound speed

The coefficients characterizing the optimized density/attenuation and sound density/speed found in eq. 3 and 6 that resulted in the best agreement between the models and the measurement are listed in Table 3. In optimizing the attenuation/density relationship, we used a simplified attenuation model that allowed us to rapidly estimate the focal heating for different scenarios. We estimated the temperature rise for every patient and sonication for 10,000 different attenuation vs. density curves. The best result we found (excluding the deviants shown in FIG. 8b and d) is shown in FIG. 10a. The relationships that were within 10% of the best curve followed a similar trajectory except for higher densities that were rarely observed in these 32 patients. Using the best relationship, the full simulations were repeated. Comparison of the focal heating between the simplified and full simulations are shown in FIG. 10b. Overall the two estimates were highly correlated, with heating estimated by the simplified model slightly higher in most patients than that predicted by the full simulations.

After the attenuation optimization, we performed a similar procedure with sound speed. FIG. 10c shows the density/sound speed relationships that best predicted the measured peak focal heating for the deviant and non-deviant sonications. FIG. 10d compares the heating

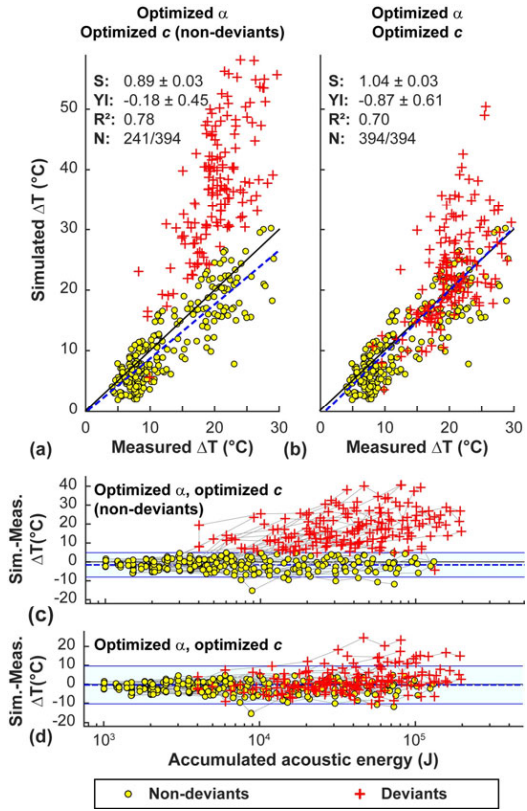

FIG. 11. Validating the density/attenuation and density/sound speed optimization in 40 additional patients. (a-b) Comparison of simulated and measured focal heating using the optimized attenuation (α) and sound speed (c) relationships shown in FIG. 10 for patients 33-72. (c-d) Plot of the difference in the simulated and measured heating as a function of the accumulated energy delivery. The results for these patients are similar to those shown in FIG. 8 for the initial 32 patients used for the optimization. (S, YI: slope, y-intercept of linear regression)

predicted using the simplified sound speed model to that predicted by the full simulation. The two were strongly correlated, but there was considerable mismatch in some patients, and on average the heating was slightly higher in the simplified model.

### E. Validating the optimization

We simulated an additional 40 ET patients (396 sonications) using the optimized relationships we found between skull density and the attenuation and sound speed. Overall the agreement between simulation and measurement in the peak temperature rise was similar for these patients for the non-deviant sonications (FIG. 11). The optimization of the density/sound speed relationship for the deviant sonications improved the prediction in many cases, but there were several cases where the simulations continued to over-predict the focal heating. Additional information about these patients is shown in Table S1 and FIG. S3. When the 72 patients were considered together (825 sonications), regression of the simulated vs. measured temperature rise at the focus yielded a slope of $1.05 \pm 0.02$, and a y-intercept of $-1.31 \pm 0.39$ ($R^2$ 0.71). The root mean square difference between simulation and measurement was $3.4 \pm 3.1°C$.

We also repeated the optimization of the density/attenuation and density/sound speed relationships in these 40 patient and for all patients (FIG. 12). The best relationships found were similar to those from patients 1-32.

### F. Effects of spatial averaging and aberration correction

We investigated the effects of spatial averaging over the imaging planes used for MRTI (FIG. 13a) and aberration correction (FIG. 13b) on the simulated temperature maps. Overall, the predicted temperature rise was 1.5 times higher before spatial averaging when the phase corrections used during treatment were used. Given that the measured heating was more diffuse than the simulations predicted, this may be overstating the effects of averaging. The simulations predicted that using the ideal phase corrections would result in improved heating 1.2 times higher than the corrections used during the treatments. The simulations also predicted that without the phase aberration corrections used during the treatments that the temperature rise would have been reduced by half on average.

### G. Predicting ablative energy

There was substantial variability among the different patients in the ultrasound exposure levels needed to achieve a temperature rise sufficient to produce a thermal lesion. To characterize this variability, we estimated the acoustic energy required in each patient to achieve an ablative thermal exposure of 55°C in MRTI (dotted vertical lines in FIG. 8a). This energy range spanned an order of magnitude; it ranged 3.7-34.6 kJ in patients 1-32 and 3.3-36.1 kJ in patients 33-72. We compared the ability of the simulations to predict this exposure level along with multiple other metrics gleaned from the CT scans in patients 1-32 (FIG. S4). Correlation between the measured and simulated energy needed to reach 55°C was significant (P<0.001; $R^2$: 0.45), but with substantial variation. Several other skull-derived metrics also showed a significant correlation. The three cases that required the highest energy were outliers in these regressions, and the high acoustic energy required to reach an ablative thermal exposure at the focus (>20 kJ) was not predicted.

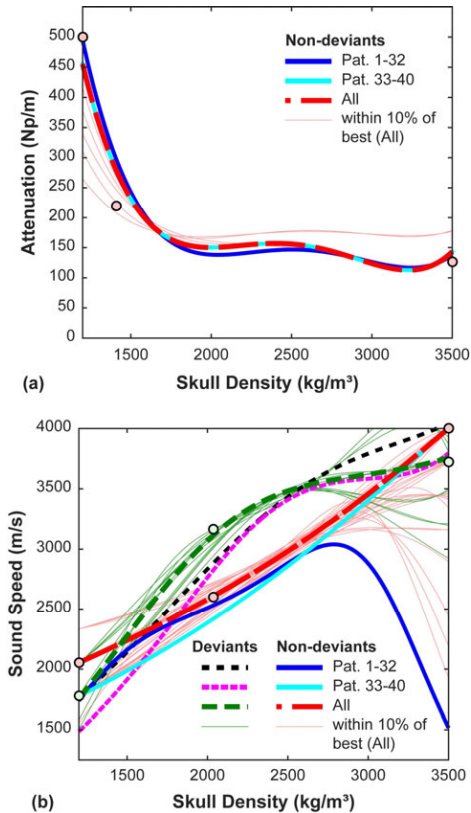

**FIG. 12.** Optimizing density/attenuation and density/sound speed relationships using results from 72 patients. The relationships found were similar for those for patients 1-32.

FIG. 14 shows color-coded maps of the skulls for patients 1-32 showing the spatial distribution of three metrics that had highly-significant correlations with the energy needed to reach 55°C: loss in pressure due to skull attenuation, skull thickness, and cortical/trabecular density ratio. The patients are ordered by the acoustic energy needed to reach 55°C. While trends are evident, there was substantial variability. Note, for example patient 19, who's skull was relatively thick and highly attenuating but required very little energy, and patient 30, who's skull was thinner and less attenuating than others that required less energy.

We also compared these skull-derived metrics for patients that did and did not have sonications with heating/energy trajectories that deviated from the predictions. The skulls of the patients that had sonications with deviations were thinner (6.8 ± 1.1 vs. 8.1 ± 1.1 mm; P<0.01) and had a smaller volume (275 ±

51 vs. 329 ± 62 cm³; P<0.05) than those where the trajectories did not deviate. The corresponding loss due to attenuation was also less (0.69 ± 0.06 vs. 0.79 ± 0.74 ± 0.05; P<0.05) in the patients with deviants. The best predictor of whether such simulation/measurement mismatch occurred was the standard deviation of the attenuation along the direction of ultrasound propagation (68.6 ± 4.9 vs. 61.3 ± 5.0 Np/m; P<0.001).

## H. Comparing "ideal" and treatment phase aberration corrections

The phases used for aberration correction determined by the device software and used in the treatments were grossly similar to the "ideal" phases predicted by the simulations in every case in patients 1-32 ($R^2$: 0.84, P<0.001; FIG. S5). We examined whether the different metrics gleaned from the CT scans were correlated with differences in the predicted ideal phase corrections and those used during treatment. Factors that had a significant correlation with the difference between the ideal and treatment phases were the number of elements with incidence angles greater than 25° (P<0.001), the standard deviations of the density, sound speed, and impedance, the density and sound speed of the outer table (P<0.05), and the difference between the internal and external incidence angles (P=0.001). The proprietary model used by the manufacturer considers shear mode transmission for incidence angles above a certain threshold, so it is not surprising that differences that were correlated with factors related to incidence angles.

## IV. DISCUSSION

This work describes our experience testing a simulation framework that enables rapid calculation of the pressure field for different transducer magnitude/phase distributions and that could be iterated using simplified models to identify relationships between the CT-derived density and the skull acoustic properties that better matched the MRTI. This approach required more upfront work to generate the individual elemental simulations, but it could be performed in parallel using a computing cluster and potentially allows for more flexibility in exploring how changing different properties affects focal heating.

Ultimately, we aim to assemble a look-up table of elemental skull segments and pressure distributions to enable us to accurately predict the three-dimensional pressure distribution in real time. This table could be based on similarity comparisons between skull segments

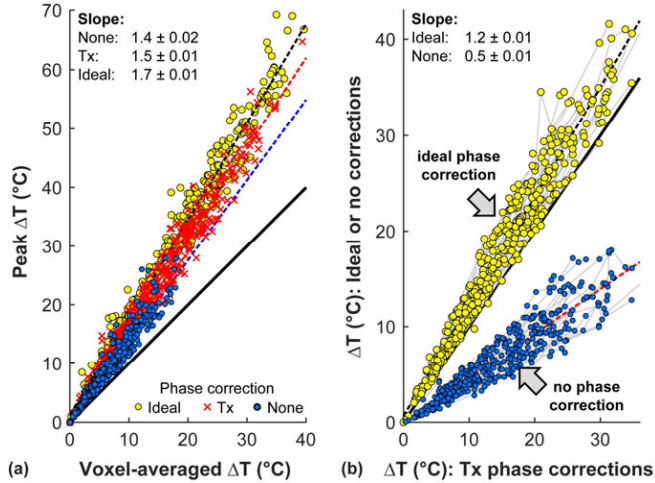

(a) Voxel-averaged ΔT (°C)  (b) ΔT (°C): Tx phase corrections

FIG. 13. Impact of spatial averaging and aberration correction on focal heating. (a) Effects of spatial averaging on the simulated peak temperature rise at the focus using the phase corrections calculated by the manufacturer, the ideal phase corrections, and no correction. The models suggest that the peak temperature rise was 1.4-1.7 times the value measured with MRTI. (b) Plot of simulated temperature rise using ideal phase corrections and no correction vs. those used during treatment (Tx). The models suggest that the correction used during the treatment improved the measured heating by a factor of 2, and that using the ideal correction predicted could yield a mean improvement of approximately 20% (dotted lines: linear regressions; solid lines: unity)

or based on derived factors such as thickness, angles, etc. Machine learning methods may be useful in developing these predictive models to relate the skull segments to the resulting pressure distributions. If one had the ability to predict the three-dimensional pressure distribution in real time, one could rapidly iterate the phase and magnitude of the transducer elements (or the position and angulation of the transducer itself) to optimize not only the peak temperature rise but also the shape and size of the focal heating. With an ability to predict the pressure amplitude in real time, one could also apply holographic methods [29] to shape the focal region to match the desired anatomy of interest. Such shaping may be more useful in TcMRgFUS applications other than thermal ablation, such as blood-brain barrier disruption [6,7], neuromodulation [30], and nonthermal ablation [31] that do not require high acoustic exposure levels, as modifying the spatial pressure field would likely reduce the peak pressure amplitude at the focus.

To achieve this goal, we need to better predict the shape of the focal heating. The heating measured with MRTI was more diffuse than the simulations predicted. This result could suggest that acoustic parameters used in the model were not correct, or it could reflect factors that were not considered here in the simulations. Transmission after shear mode conversion in elements with oblique incidence angles, for example, was not accounted for in these simulations. While the high shear attenuation coefficient will limit this transmission, studies in cadaver skulls suggest that transmission of up to 17-23% of the power can be transmitted via a shear wave for incidence angles greater than 30° [32]. Complex reflections, such as those that occur from the face of the transducer and those that occurred outside of the element-wise simulation volumes may have also contributed, and probably to a small effect [33], nonlinear acoustic propagation. These components of the transmitted acoustic wave were not taken into account in the aberration correction, perhaps leading to diffuse heating at and around the focal region in the measurements that was not present in the simulations. This diffuse heating may explain the slower temperature decay after the sonications (FIG. 8b). A faster decay in the simulation might also suggest that the thermal properties and the perfusion coefficients used for brain were incorrect.

Other than the diffuse heating, the simulations did a good job overall in predicting the relative peak temperature rise and orientation/obliquity of the focal region. However, it is possible that the parameterization of the CT scans and the acoustic and thermal properties need refinement. For example, based on earlier works [14,15], we used a simple linear extrapolation to estimate skull density from Hounsfield units, and the Hounsfield units were estimated directly using information from the DICOM headers. Others have suggested potentially more accurate approaches to estimate skull density from CT scans [16,34], and have shown that scans from CT scanners from different vendors and with different reconstruction kernels can yield different results [35]. Here, the CT scans were from a variety of vendors, but we did not see any obvious effects on the simulations.

We began this project using the experimentally-derived skull density/sound speed and density/attenuation relationships found by Pichardo et al. [15]. While we did observe good agreement between the measurements and the simulations in some patients, in most cases the peak temperature was under-predicted by the numerical model for sonications at relatively low exposure levels. Furthermore, in many patients the relationship between acoustic energy and temperature rise followed a similar trajectory in the initial sonications

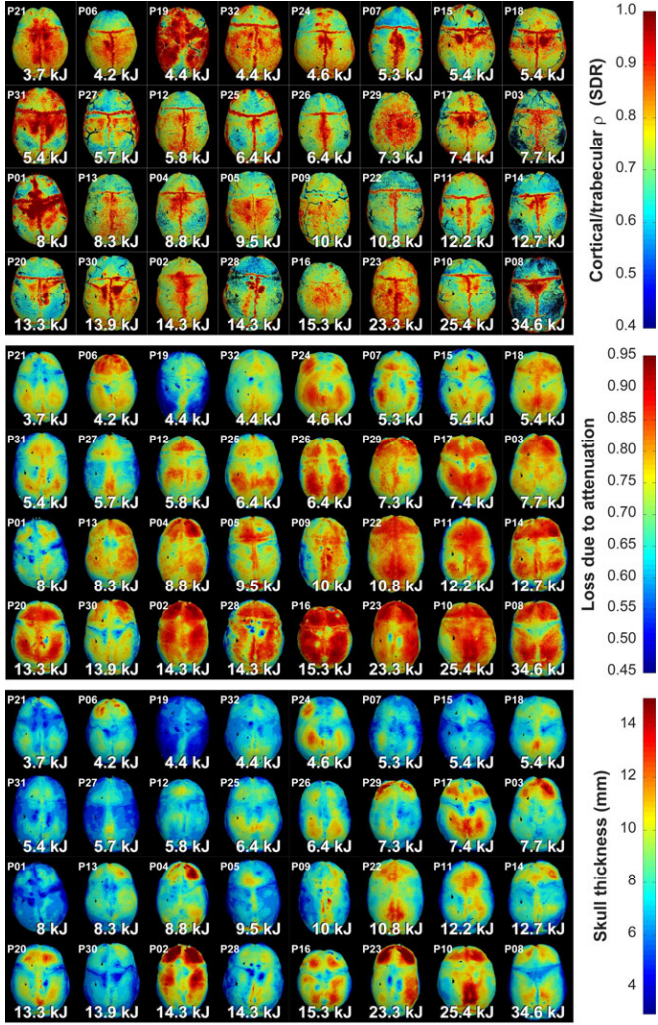

FIG. 14. Skull-derived metrics and treatment efficiency. Maps of the outer skull surface color-coded based on different metrics derived from the CT scans are shown for 32 patients. The patients are sorted by the acoustic energy needed to reach a focal temperature of 55°C, which covered an order of magnitude. A general trend towards increasing acoustic energies as the loss due to attenuation, skull thickness, and attenuation coefficient of the diploe increased and as the trabecular/cortical density ratio decreased is evident. Clear outliers for each metric are also evident.

but appeared to deviate after a certain amount of acoustic energy was applied.

Based on these results, as well as clinical experience showing that increasing acoustic energy often produces diminishing increases in heating as the treatment progresses [11] and that has observed skull damage in some patients after treatment at high energies [36], we hypothesized that (1) the relationship between attenuation and skull density found by Pichardo et al.

was not optimal, and (2) the deviation could represent an irreversible change in skull acoustic properties. We attempted to find a density/attenuation relationship that better predicted the MRTI measurements for sonications delivered before the skull changes. This process improved the predictive ability of the simulations substantially, but there were still patients where the agreement was poor. It is possible that there is no single relationship between density and attenuation, as we do not capture the microstructure of the bone below the resolution of the CT scan, which likely affects the scattering of the acoustic field [37,38]. More work is needed to understand how well we can use clinical imaging to predict the attenuation for an individual patient.

We also explored whether we could find a skull density/sound speed relationship that could be used after a (presumed) change in skull properties. While we were able to identify a relationship that improved the predictive ability of the model after the simulated heating deviated from the measurements, there was still substantial variability. Validation of these relationships in 40 additional patients revealed similar agreement between predicted and measured heating overall. However, the density/sound speed relationship found in patients 1-32 after the presumed skull changes had even more variability in patients 33-72. Finding such variability is probably not surprising, since heat-induced changes to the skull will not be spatially uniform or simply binary. It may not even be correct to assume that the sound speed is the dominant factor that changes during the treatment.

Furthermore, our interpretation of the results that the skull acoustic properties changed may not be correct for every patient. Note, for example, patients 27 and 30 in FIG. 8a, where the predicted heating with the optimized density/attenuation relationship was substantially higher than the measurements even though the previously-delivered energy was low. Similar findings were observed in several treatments of the 40 additional patients (FIG. 11). These results might suggest that the deviations in temperature/energy were not due changes in skull properties from previously-applied sonications, but perhaps instead a dynamic change in acoustic properties with heating. It could also be that the sonications in these patients should not have been included as "deviants" and that our optimized density/attenuation relationship needs further refinement. Since we do not have heating measurements at intermediate energies in these patients, we cannot answer this question. Future studies with additional patients are needed to understand cases like these.

We have demonstrated how one could use element-wise simulations to improve the relationships between acoustic attenuation and sound speed. Here, we used simplified models that could be rapidly applied to previously-obtained simulations to explore the effects of thousands of different relationships on resulting focal heating. While these simplified models did a good job in predicting the temperature at the focus for the full simulation, a better approach would be to explore these relationships to match not only the peak temperature but also the shape and size of the focal heating. Perhaps it would also be better to optimize over multiple acoustic and thermal parameters simultaneously. Finally, we need to evaluate whether the relationships between density and acoustic properties found here can improve the focusing in future TcMRgFUS treatments.

## A. Predictors of treatment efficacy

We observed an order of magnitude range in energy required to achieve an effective level of heating. It would be useful to predict this energy before treatment to screen potential TcMRgFUS patients. Currently, the device manufacturer uses the "skull density ratio", which is the ratio in density between the diploic trabecular and dense cortical bone [10,39] to predict which patients will require high acoustic energies to achieve focal temperatures sufficient to induce a thermal lesion. We examined different skull-derived parameters to see how they compare in this prediction. The manufacturer-derived "skull density ratio" as well as our own calculation of this ratio were both predictive of the energy needed to reach 55°C ($R^2$: 0.30, 0.35, respectively; $P<0.01$), but a number of other parameters, such as the skull thickness and the expected loss due to attenuation were also similarly predictive. However, there was significant variability and there were outliers that were not predicted by simple skull measurements. The simulations did a better job in predicting the acoustic energy needed to reach 55°C ($R^2$: 0.45; $P<0.001$) than any single factor, but there is still considerable room for improvement.

## B. Conclusions

This study demonstrates the feasibility of using an element-wise approach to simulate TcMRgFUS thermal ablation and predict the shape and magnitude of the focal heating in multiple orientations. While there was significant variability, the numeric model predicted the relative shape and focal temperature rise on average. Deviations between the measured and simulated heating were observed in many patients over the course of the treatments, perhaps reflecting a change in skull properties. We also demonstrated how this approach could be used to optimize the relationship between CT-derived density and the acoustic attenuation and sound speed. We found initial estimates of these relationships based on 32 patient treatments, including an estimated density/sound speed relationship that could be used after the model and the measurements deviated. The optimization was validated in 40 additional patients. Future work will expand on this study to refine these optimized relationships.

## ACKNOWLEDGEMENTS

This work was supported by NIH grant R01EB025205. The authors thank Greg Clement, Alex Hughes, Samuel Pichardo, Urvi Vyas, and Brad Treeby for their helpful advice, and the engineers at InSightec for their help in extracting the transducer and sonication parameters for these treatments.

Portions of this research were conducted on the O2 High Performance Compute Cluster, supported by the Research Computing Group at Harvard Medical School. See http://rc.hms.harvard.edu for more information.

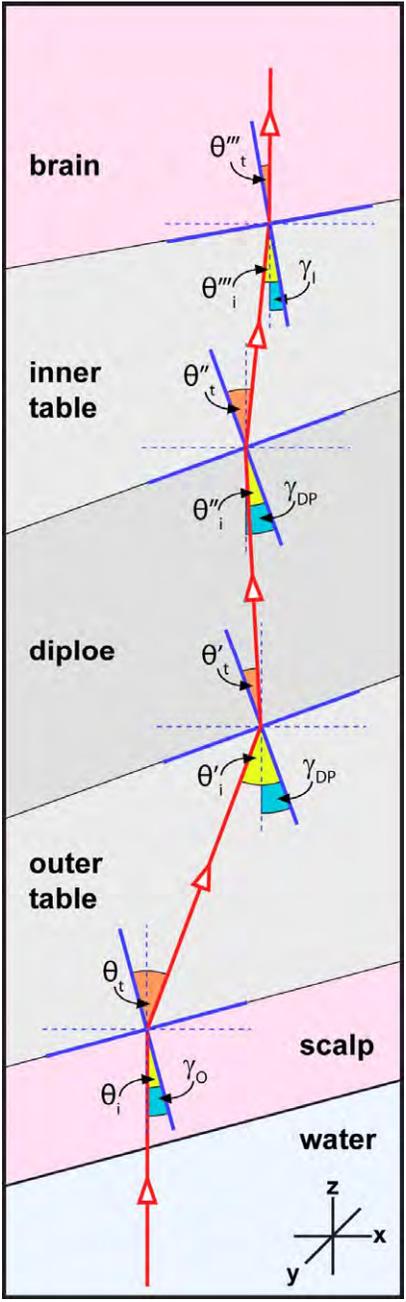

FIG. S1. Definitions of the angles used in the simplified model used to optimize the density/sound speed relationship. The incidence angles between the direction of ultrasound propagation ($z$) for each transducer element and the outer and inner skull surface and the diploe ($\gamma_O$, $\gamma_I$, and $\gamma_{DP}$) was obtained for each 44×44 $x,y$ coordinate. We assumed that the incidence angles between the outer and inner tables and the diploe were the same.

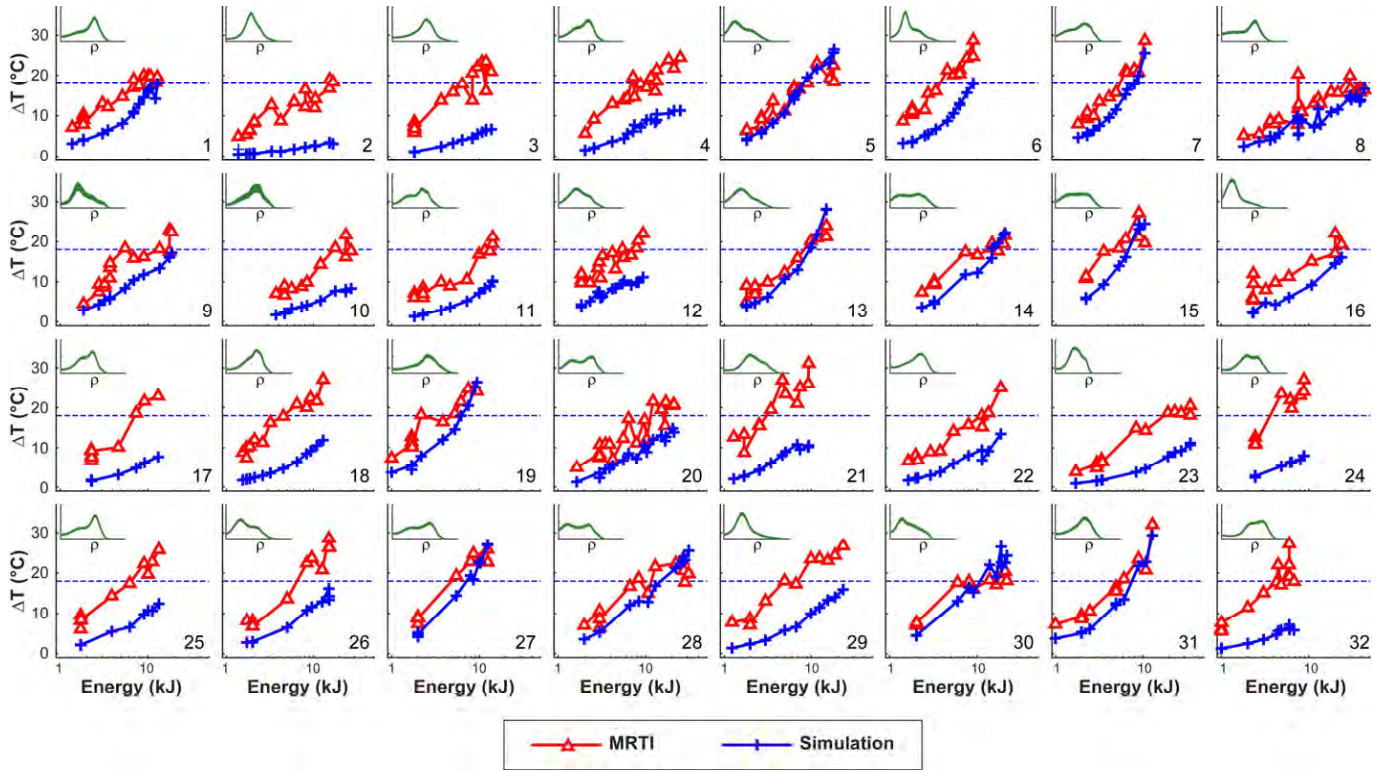

FIG. S2. Measured and simulated focal heating as a function of the applied acoustic energy for 32 patients using the attenuation/density and sound speed/density relationships found by Pichardo et al. [15]. The horizontal dotted lines indicate an absolute temperature of 55°C, a rough estimate for thermal necrosis.

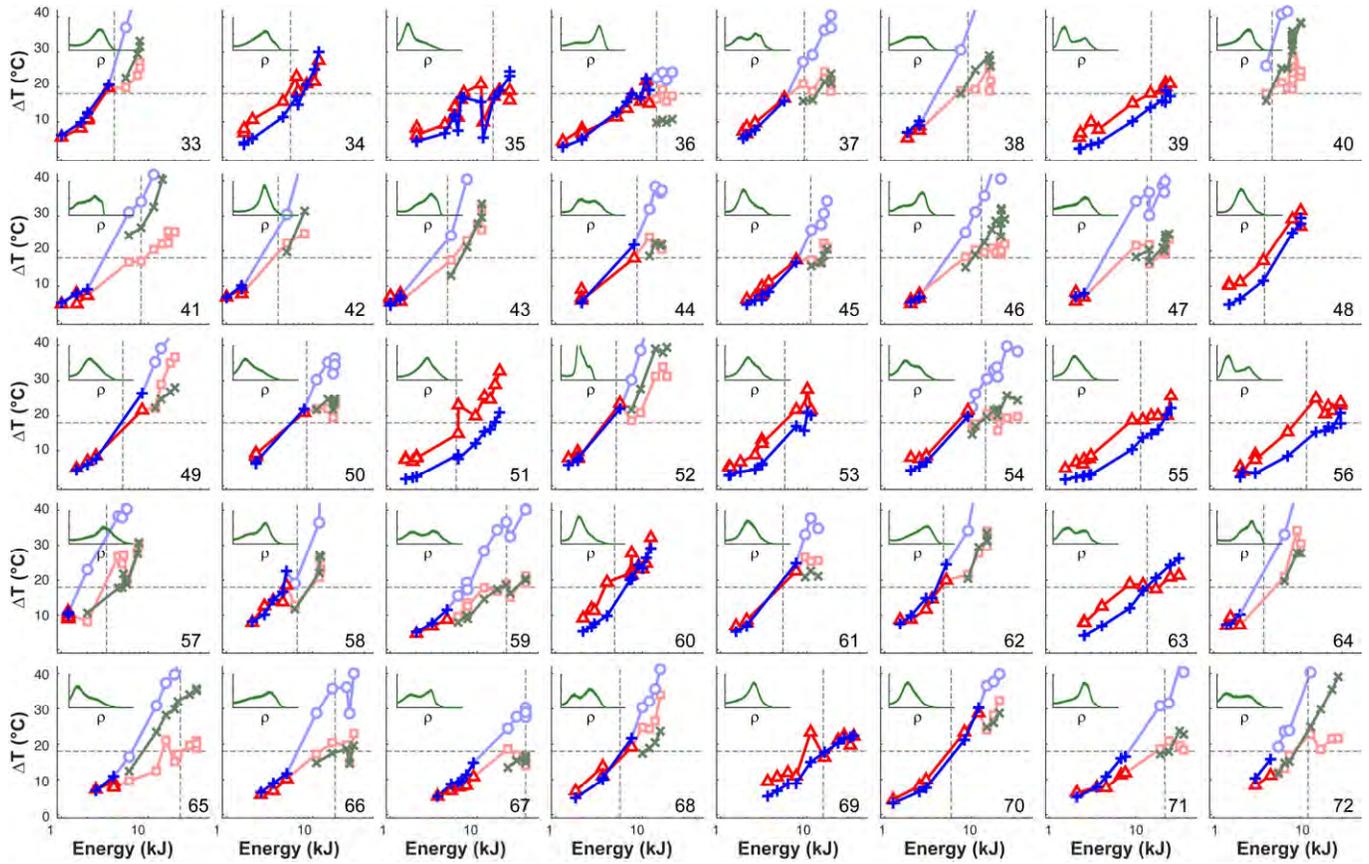

FIG. S3. Measured and simulated focal heating as a function of the applied acoustic energy for 40 patients using the attenuation/density and sound speed/density relationships optimized in patients 1-32. The horizontal dotted lines indicate an absolute temperature of 55°C, a rough estimate for thermal necrosis. The insets show histograms of skull density (x-axis: 1290-3500 kg/m³)

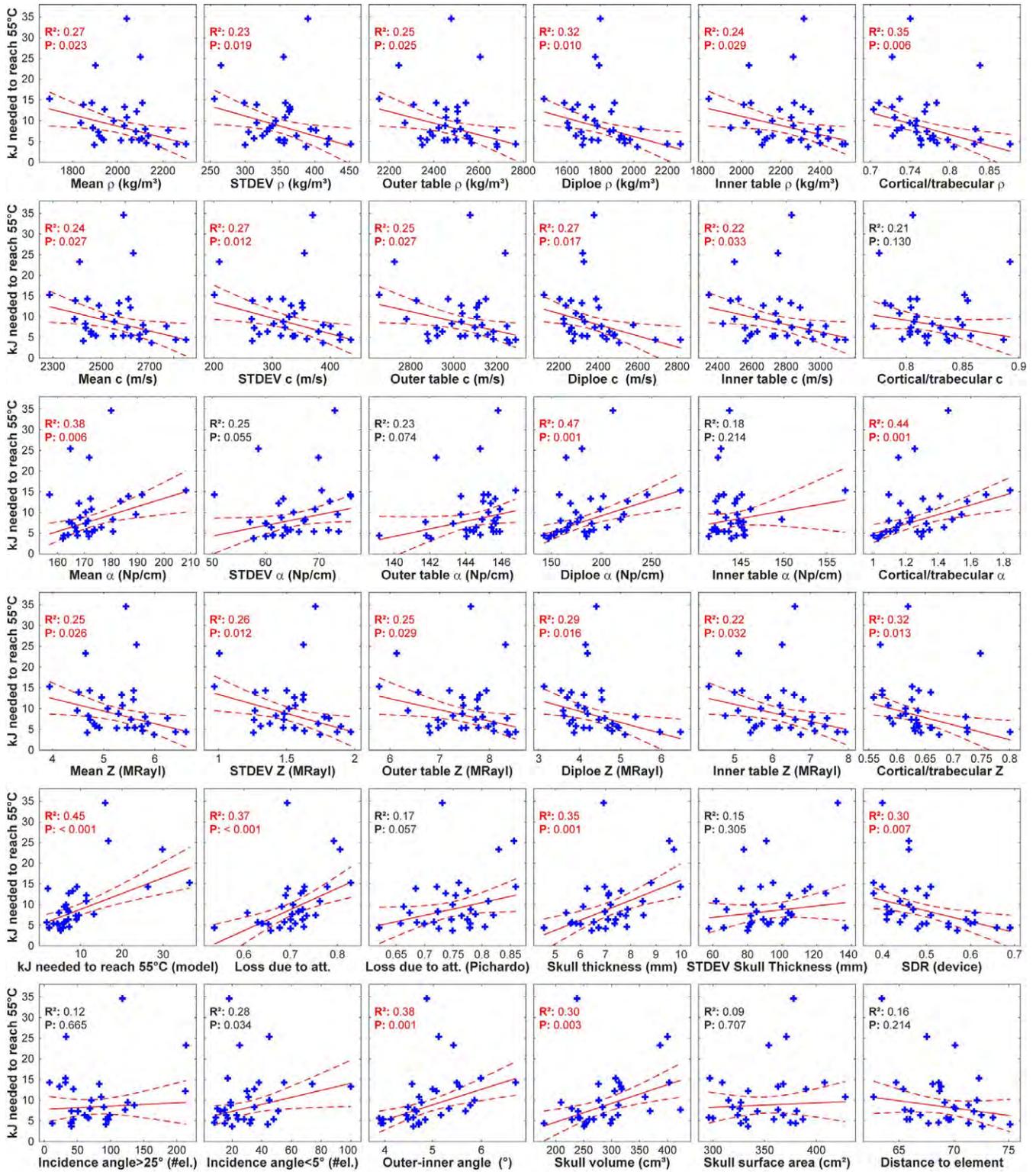

FIG. S4. Investigating predictors of the acoustic energy needed to achieve a focal temperature of 55°C. Features were extracted on a per-element basis from the CT scans. Results of robust linear regression are noted. In most cases, the three patients that required the highest energy were outliers. Results from the simulations are also included as well as the "skull density ratio" (SDR) provided by the device manufacturer.

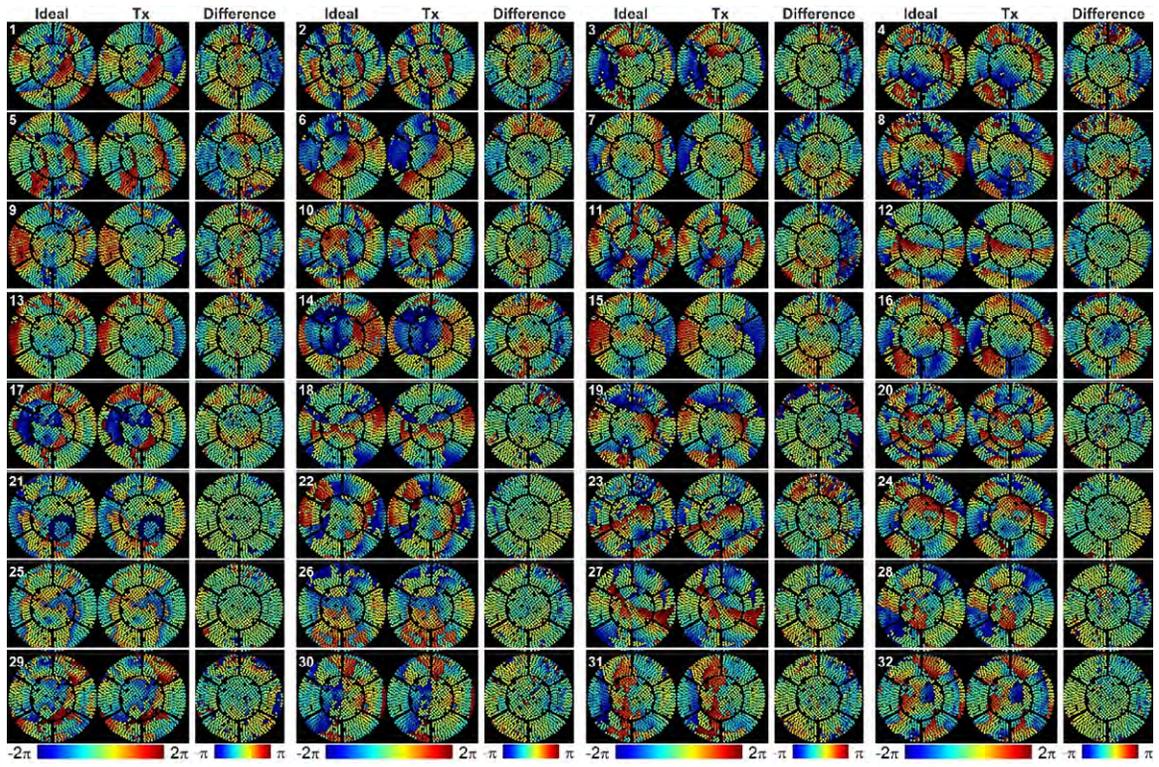

FIG. S5. Predicted vs. treatment phase corrections. Maps of the 1024-element transducer array showing the phase offsets used for aberration correction during the treatment, "ideal" corrections predicted by the simulations, and their difference, for 32 patients. The simulation produced phase offsets that were similar to those used in the treatments in every patient for most of the elements.

Table S1. Patient, treatment, and CT information for the 40 patients in the validation study.

| Patient | | | Treatment | | | | | | CT Scan | | | |
|---|---|---|---|---|---|---|---|---|---|---|---|---|
| | | | | | | Acoustic energy (kJ) | | | | | | |
| N | Age | SDR | # son. | Power (W) | Duration (s) | Max. | Total | To reach 55°C | Vendor | Kernel | keV | Voxel (mm) |
| 33 | 73M | 0.75 | 11 | 94-744 | 11-13 | 8.0 | 43.4 | 4.1 | GE | BONE+ | 120 | 0.49×0.49×0.63 |
| 34 | 67F | 0.63 | 9 | 145-986 | 12-13 | 11.8 | 54.0 | 5.6 | GE | BONE+ | 120 | 0.43×0.43×0.63 |
| 35 | 85F | 0.41 | 16 | 185-987 | 10-36 | 23.7 | 144.6 | 15.2 | SIEMENS | H60s | 120 | 0.43×0.43×1.00 |
| 36 | 72M | 0.58 | 14 | 140-893 | 10-32 | 22.1 | 131.0 | 14.6 | GE | BONE+ | 120 | 0.49×0.49×0.63 |
| 37 | 68M | 0.46 | 10 | 184-1018 | 12-20 | 19.1 | 88.8 | 9.4 | SIEMENS | H60s | 130 | 0.44×0.44×1.00 |
| 38 | 81M | 0.41 | 8 | 184-1112 | 12-18 | 16.6 | 74.3 | 9.1 | SIEMENS | H60s | 120 | 0.41×0.41×1.00 |
| 39 | 72F | 0.40 | 10 | 184-1241 | 12-23 | 25.0 | 124.3 | 15.0 | SIEMENS | Hr59h\2 | 120 | 0.45×0.45×1.00 |
| 40 | 67M | 0.55 | 15 | 369-840 | 12-13 | 10.1 | 117.6 | 4.7 | SIEMENS | H60s | 120 | 0.47×0.47×1.00 |
| 41 | 74M | 0.53 | 11 | 91-1118 | 12-19 | 20.2 | 100.8 | 8.3 | Philips | C | 120 | 0.49×0.49×1.00 |
| 42 | 76F | 0.68 | 5 | 93-740 | 12-12 | 8.1 | 17.2 | 4.0 | SIEMENS | H60s | 120 | 0.43×0.43×1.00 |
| 43 | 66M | 0.65 | 12 | 91-1068 | 11-13 | 11.3 | 53.2 | 4.5 | GE | BONE+ | 140 | 0.49×0.49×0.63 |
| 44 | 75M | 0.49 | 8 | 186-1114 | 12-16 | 16.8 | 73.2 | 8.7 | SIEMENS | H60s | 100 | 0.50×0.50×1.00 |
| 45 | 60F | 0.55 | 10 | 187-1131 | 12-18 | 17.2 | 80.7 | 11.1 | GE | BONE+ | 120 | 0.48×0.48×1.00 |
| 46 | 78M | 0.44 | 17 | 184-1297 | 12-23 | 24.2 | 216.4 | 13.0 | SIEMENS | H60s | 120 | 0.49×0.49×1.00 |
| 47 | 72M | 0.42 | 12 | 186-1128 | 12-28 | 24.1 | 176.8 | 13.5 | SIEMENS | H60s | 120 | 0.50×0.50×1.00 |
| 48 | 68F | 0.56 | 7 | 140-745 | 12-14 | 10.0 | 36.9 | 3.8 | SIEMENS | H60s | 120 | 0.46×0.46×1.00 |
| 49 | 73F | 0.50 | 8 | 139-1130 | 12-19 | 20.3 | 79.0 | 5.1 | SIEMENS | H60s | 120 | 0.41×0.41×1.00 |
| 50 | 89F | 0.38 | 10 | 231-1122 | 10-22 | 18.1 | 96.5 | 8.6 | SIEMENS | H60s | 120 | 0.41×0.41×1.00 |
| 51 | 70F | 0.51 | 12 | 138-1126 | 8-17 | 18.1 | 90.9 | 5.7 | TOSHIBA | FC30 | 120 | 0.40×0.40×1.00 |
| 52 | 68F | 0.51 | 10 | 138-1109 | 11-18 | 19.2 | 79.9 | 5.0 | Philips | C | 120 | 0.45×0.45×1.00 |
| 53 | 75F | 0.51 | 10 | 138-1116 | 10-15 | 11.2 | 51.0 | 5.6 | SIEMENS | H60s | 120 | 0.45×0.45×1.00 |
| 54 | 84M | 0.44 | 12 | 186-1124 | 12-37 | 33.9 | 170.2 | 14.5 | SIEMENS | H60s | 120 | 0.46×0.46×1.00 |
| 55 | 73F | 0.53 | 12 | 137-1285 | 12-21 | 25.5 | 117.1 | 11.3 | SIEMENS | H60s | 120 | 0.47×0.47×1.00 |
| 56 | 73F | 0.40 | 12 | 184-1134 | 12-28 | 28.9 | 154.2 | 11.7 | SIEMENS | Hr59h\2 | 120 | 0.45×0.45×1.00 |
| 57 | 76M | 0.67 | 12 | 137-782 | 10-12 | 7.8 | 43.9 | 3.3 | TOSHIBA | FC30 | 120 | 0.47×0.47×1.00 |
| 58 | 68M | 0.48 | 11 | 149-846 | 9-18 | 12.4 | 60.5 | 12.4 | SIEMENS | H60s | 120 | 0.47×0.47×1.00 |
| 59 | 79M | 0.44 | 12 | 187-1167 | 11-32 | 36.1 | 177.5 | 21.7 | SIEMENS | H60s | 100 | 0.46×0.46×1.00 |
| 60 | 79F | 0.53 | 11 | 190-1139 | 11-12 | 12.5 | 76.5 | 4.8 | SIEMENS | H60s | 120 | 0.46×0.46×1.00 |
| 61 | 66F | 0.55 | 7 | 140-1039 | 11-19 | 13.6 | 47.3 | 5.3 | SIEMENS | H60s | 120 | 0.48×0.48×1.00 |
| 62 | 65M | 0.62 | 10 | 141-1146 | 11-14 | 15.1 | 81.7 | 4.8 | GE | BONE+ | 120 | 0.49×0.49×0.63 |
| 63 | 87M | 0.41 | 8 | 235-1172 | 11-30 | 31.5 | 103.3 | 13.5 | SIEMENS | H60s | 120 | 0.47×0.47×1.00 |
| 64 | 83M | 0.60 | 6 | 141-938 | 11-12 | 10.0 | 30.6 | 3.8 | Philips | D | 120 | 0.46×0.46×1.00 |
| 65 | 92F | 0.38 | 15 | 239-1144 | 7-36 | 36.1 | 236.0 | 23.4 | SIEMENS | H60s | 120 | 0.45×0.45×1.00 |
| 66 | 87M | 0.46 | 8 | 236-1126 | 11-44 | 30.0 | 120.8 | 18.2 | GE | BONE+ | 120 | 0.49×0.49×0.63 |
| 67 | 74M | 0.43 | 12 | 331-1101 | 11-34 | 36.1 | 232.7 | 36.1 | SIEMENS | H60s | 120 | 0.45×0.45×1.00 |
| 68 | 73M | 0.53 | 8 | 190-1122 | 11-16 | 16.1 | 68.5 | 5.5 | SIEMENS | Hr59h\2 | 120 | 0.53×0.53×1.00 |
| 69 | 79F | 0.44 | 10 | 332-1043 | 11-40 | 35.0 | 165.9 | 15.5 | SIEMENS | H60s | 120 | 0.41×0.41×1.00 |
| 70 | 89F | 0.51 | 8 | 140-1129 | 12-20 | 21.0 | 81.2 | 5.8 | SIEMENS | H60s | 120 | 0.45×0.45×1.00 |
| 71 | 89M | 0.51 | 9 | 191-1145 | 12-37 | 35.8 | 135.4 | 21.6 | SIEMENS | H60s | 120 | 0.47×0.47×1.00 |
| 72 | 69M | 0.43 | 9 | 281-1136 | 11-25 | 27.0 | 107.2 | 12.2 | SIEMENS | H60s | 120 | 0.47×0.47×1.00 |

SDR: "skull density ratio"

## Supplemental Methods

In the exploration the ability of different skull-derived factors to predict the energy needed to reach 55°C and to estimate the transmission coefficient in the simplified sound speed model, we used the following formulae. The simulation (*xyz*) space consisted of 44×44×492 elements. The density of the skull was interpolated into this space for each element. For each *x* and *y* coordinate of these elemental volumes, we calculated the following factors:

Mean skull density, sound speed, attenuation, impedance:

$$\overline{\rho_z} = \frac{1}{z_b - z_a} \sum_{z=z_a}^{z=z_b} \rho(z), \quad \overline{c_z} = \frac{1}{z_b - z_a} \sum_{z=z_a}^{z=z_b} c(\rho(z)), \quad \overline{\alpha_z} = \frac{1}{z_b - z_a} \sum_{z=z_a}^{z=z_b} \alpha(\rho(z)), \quad \overline{Z_z}$$

$$= \frac{1}{z_b - z_a} \sum_{z=z_a}^{z=z_b} \rho(z) \cdot c(\rho(z))$$

Where *a* and *b* are the coordinates of the outer and inner surface, respectively. We also calculated the standard deviation of the acoustic properties for the points between $z_a$ and $z_b$.

Skull density, sound speed, attenuation at the outer table:

$$\overline{\rho_{OT}} = \frac{1}{3} \sum_{z=z_{OT}-1}^{z=z_{OT}+1} \rho(z), \qquad \overline{c_{OT}} = \frac{1}{3} \sum_{z=z_{OT}-1}^{z=z_{OT}+1} c(\rho(z)), \qquad \overline{\alpha_{OT}} = \frac{1}{3} \sum_{z=z_{OT}-1}^{z=z_{OT}+1} \alpha(\rho(z)),$$

$$\overline{Z_{OT}} = \frac{1}{3} \sum_{z=z_{OT}-1}^{z=z_{OT}+1} \rho(z) \cdot c(\rho(z))$$

These values at the inner table and the diploe were found in the same way.

Loss due to skull attenuation:

$$L = \exp\left( -\sum_{z=z_a}^{z=z_b} \alpha(\rho(z)) \cdot \Delta z \right)$$

Trabecular/cortical ratio:

$$\rho_{ratio} = \frac{0.5 \cdot (\overline{\rho_{IT}} + \overline{\rho_{OT}})}{\overline{\rho_{DP}}} \quad c_{ratio} = \frac{0.5 \cdot (\overline{c_{IT}} + \overline{c_{OT}})}{\overline{c_{DP}}} \quad \alpha_{ratio} = \frac{0.5 \cdot (\overline{\alpha_{IT}} + \overline{\alpha_{OT}})}{\overline{\alpha_{DP}}} \quad Z_{ratio}$$

$$= \frac{0.5 \cdot (\overline{Z_{IT}} + \overline{Z_{OT}})}{\overline{Z_{DP}}}$$

The mean value of each metric was calculated for each transducer element. Results presented are the means or standard deviations over all the elements for each patient.